\PassOptionsToPackage{hypertexnames=false}{hyperref}

\documentclass[12pt, a4paper,reqno]{amsart}
\usepackage{amsmath,amssymb,amsthm}
\usepackage{enumitem}
\usepackage{doi,url}
\usepackage{dsfont}
\usepackage{mathabx}
\usepackage[dvipsnames]{xcolor}
\usepackage{dutchcal} 
\usepackage{tikz}
\usepackage{tkz-euclide}
\usetikzlibrary{shapes.geometric, calc}
\usepackage{float}

\numberwithin{equation}{section}
\newtheorem{theorem}{Theorem}[section]
\newtheorem{lemma}[theorem]{Lemma}
\newtheorem{proposition}[theorem]{Proposition}
\newtheorem{corollary}[theorem]{Corollary}

\theoremstyle{definition}
\newtheorem{remark}[theorem]{Remark}

\newcommand{\set}[1]{\left\{ #1 \right\}}

\newcommand{\N}{{\mathbb N}}
\newcommand{\Z}{{\mathbb Z}}
\newcommand{\C}{{\mathbb C}}
\DeclareMathOperator{\supp}{supp}

\newcommand{\norm}[1]{\left\lVert #1 \right\rVert}
\newcommand{\pa}[1]{\left( #1 \right)}
\DeclareMathOperator{\tr}{tr}
\DeclareMathOperator{\diam}{diam}

\newcommand{\abs}[1]{\left\lvert #1 \right\rvert}

\newcommand\be{\begin{equation}\begin{aligned}}
\newcommand\ee{\end{aligned}\end{equation}}
\newcommand{\R}{\mathbb R}
\title{Lieb--Robinson bounds for Bose--Hubbard Hamiltonians via path expansions}
\author{Alexander Elgart}
\address{Department of Mathematics, Virginia Tech, Blacksburg, VA 24061, USA}
\email{aelgart@vt.edu}
\author{Marius Lemm}
\address{Department of Mathematics, University of T\"ubingen, 72076 T\"ubingen, Germany}
\email{marius.lemm@uni-tuebingen.de}
\date{}

\begin{document}

\begin{abstract}
We study the transport properties of Bose--Hubbard-type Hamiltonians. Kuwahara, Vu, and Saito proved sharp bounds on (i) particle transport and (ii) the Lieb-Robinson velocity for initial states with controlled boson density.
We provide simpler and shorter proofs for both results using a new set of methods. For (i), boson transport, we use a class of weighted extended particle number operators that directly enables closing the associated differential inequalities. For (ii), the Lieb-Robinson bounds, we develop a bosonic version of the Chen--Lucas path expansion technique and recover the sharp velocity bound $t^{d-1}$ up to logarithmic corrections, where $d$ denotes the lattice dimension.
We also extend the transport bounds (i) to long-range hopping with a power-law exponent $\alpha>d$. The range $\alpha\in (d,d+1)$ corresponds to an algebraic light cone for boson transport and has not been previously covered.    
\end{abstract}
\maketitle

\section{Introduction}
A fundamental  result in the mathematical theory of quantum spin systems, particularly those characterized by short-range interactions, is the finite speed of information propagation. It establishes that the Heisenberg time evolution $A(t)$ of an initial local observable $A$, which is originally supported at the origin, remains essentially supported within a ball of radius $v|t|$ centered at the origin. Any signal or influence outside of this effective light cone is constrained by corrections that are exponentially small in terms of the distance beyond the radius $v|t|$. The parameter $v$ is a finite constant known as the Lieb--Robinson velocity, a term derived from its origin in the celebrated Lieb--Robinson bound \cite{LR72}.

The mathematical basis for the finiteness of $v$ in spin systems is fundamentally tied to the structural property that local terms within a spin Hamiltonian are uniformly bounded in norm. Consequently, the local energy scales within these models are also uniformly bounded, providing a natural constraint on the speed of dynamics. In a broader context, it is generally expected that the Lieb--Robinson velocity $v$, which is a many-body analog of the group velocity, scales linearly in the local energy, a relationship that is corroborated by various exactly solvable models. However, this expectation encounters significant theoretical hurdles when applied to  bosonic lattice models. These systems are characterized by unbounded local energies, making the establishment of a rigorous and meaningful speed of information propagation a  challenge.  In particular, operator norm bounds, which are sensitive to arbitrarily locally concentrated initial states, are no longer expected outside of special situation such as perturbations of harmonic systems \cite{NRSS09} or dissipative systems \cite{lemm2026dissipation}. The problem with operator norms is that they are sensitive to initial states where a macroscopic number of all $N$ bosons is concentrated in the region of interest, which leads to a speed of information propagation proportional to $N$; see \cite[Supplemental Material, Section~I]{faupin2025macroscopic} and Figure 1 in \cite{lemm2026dissipation}.
Even for better behaved initial states, there exist one-dimensional, translation-invariant bosonic Hamiltonians with exponentially fast transport \cite{eisert2009supersonic}, meaning that the distance covered grows exponentially in time. 

Here, we focus on the paradigmatic and physically relevant class of bosonic lattice models known as \textit{Bose-Hubbard type Hamiltonians} which are of the form
\[
H=\sum_{x,y\in G} J_{\langle x,y\rangle} (b_x^*b_y+h.c.) +\sum_{Z\subset G} V_Z(\{n_x\}_{x\in Z})
\]
where $G$ is a finite subgraph of $\mathbb Z^d$. 

Propagation bounds for such systems have been an active topic in recent years \cite{schuch2011information,kuwahara2021lieb,yin2022finite,faupin2022maximal,FLS2022lieb,LRSZ2023,LRZ2023,Kuwahara2024,van2024optimal,LRZ2025,kuwahara2025ballistic,kuwahara2026symmetry,li2026macroscopic} with far-ranging applications spanning from Hamiltonian learning \cite{mobus2023dissipation} to a proof that Bose-Hubbard dynamics on an increasing sequence of volumes converge to an automorphism in the thermodynamic limit \cite{deuchert2025dynamics}.

In the breakthrough paper \cite{Kuwahara2024}, Kuwahara, Van Vu, and Saito  obtained that the Lieb–Robinson velocity can be bounded by $v\sim|t|^{d-1}$, subject to certain logarithmic corrections, for Bose-Hubbard Hamiltonians with finite-range interactions. This result holds for the physically most relevant class of initial states of controlled density and suitable local observables. A review and a simplified proof of the weaker bound $v\sim|t|^{d}$ for bounded-density initial states are given in \cite{lemm2026lieb}.
The time dependence $v\sim|t|^{d-1}$ is  optimal for general time-dependent Hamiltonians, but can be improved for Hamiltonians with strong repulsion \cite{kuwahara2026symmetry}.

In the present work, we consider families of time-dependent bosonic Hamiltonians $H(t)$, containing the class of models considered in \cite{Kuwahara2024}, and establish bounds om (i), particle transport, and  (ii), Lieb–Robinson velocities.  Our results  are very similar to those of Kuwahara, Van Vu, and Saito, but our proofs use  different techniques and are, we believe, simpler. For (i), we use a weighted infinitely extended particle number operator. In contrast to  arguments based on compactly supported number operators such as \cite{schuch2011information,FLS2022lieb,Kuwahara2024,LRZ2023,LRZ2025}, this allows us to close the relevant differential inequalities directly, which means fewer iterations are needed within the proofs. These weights were also independently introduced in the recent work \cite{rico2025instantaneous} to establish a Sobolev space type framework for dissipative Bose-Hubbard systems.   For (ii), we use a path expansion technique that goes back to Chen--Lucas \cite{chen2021operator} in which the Heisenberg dynamics of a local observable is carefully expanded into terms that effectively grow the support, while suitably ``renormalizing'' the ones that do not. To avoid confusion, we emphasize that this path-expansion technique is different from traditional proofs of Lieb--Robinson bounds \cite{hastings2006spectral,nachtergaele2006lieb} which first use bounds and then sum these over paths. Instead, the Chen--Lucas technique first  derives an identity that expresses the Heisenberg dynamics exactly as a sum over very carefully chosen paths and then the bounds are comparatively easier.
 It was recently realized in work of McDonough, Lucas, and the second author that the ``renormalization'' of the path expansion can be performed much easier by an iterative Dyson expansion \cite[Prop.\ A.12]{MLL} (see also \cite{rubiliani2026lieb}) and we use this procedure here for the first time in a bosonic context. We would like to also mention the work of Yin--Lucas \cite{yin2022finite} which  used a different bosonic path expansion in a weighted operator space to derive Lieb-Robinsion bounds for one-dimensional Bose--Hubbard systems starting from the initial state $e^{-\mu N}$. Our approach here is rather a bosonic version of \cite{chen2021operator} than a higher-dimensional version of \cite{yin2022finite}.
 
 Our new bosonic path expansion technique has the advantage that it informs where exactly the particle truncations have to be implemented. While it is still somewhat technical, it is the only other technique compared to the one in \cite{Kuwahara2024} which gives $v\sim t^{d-1}$ and the proof is considerably shorter and simpler. 
Another advantage of the path expansion technique is that it extends seamlessly to Markovian open dynamics described by Lindbladian generators \cite{rubiliani2026lieb}. It is thus a natural question if our proof can be extended to a broad class of open Bose--Hubbard systems with general that we plan to revisit in future work.

Finally, in Appendix \ref{app:A_long_range}, we  discuss how the weighted particle number operator technique to prove  can be extended to prove particle propagation bounds for Bose-Hubbard models with \textit{long-range} hopping, providing an alternative to the elaborate renormalized multiscale ASTLO machinery developed in \cite{LRZ2023,LRZ2025}. Our new approach also allows us to cover the subtle range of $\alpha\in (d,d+1)$ for the first time, where particle propagation follows an algebraic (superlinear) light cone.

\section{Setup and main results}
Consider a finite subgraph $G=(\mathcal V,\mathcal E)$ of $\mathbb{Z}^d$. We denote edges by $\langle x,y\rangle\in \mathcal E$ and we write $d(x,y)$ for the graph distance on $G$.

The bosonic Fock space $\mathcal{F}$ over $\mathcal V$ is the direct sum 
\be
\mathcal{F} = \bigoplus_{N=0}^{\infty} \ell^2_s(\mathcal V^N).
\ee
Here, $\ell^2_s(\mathcal V^N)$ represents the Hilbert space of permutation-symmetric square-summable sequences on $\mathcal V^N$. By convention, $\ell^2_s(\mathcal V^0)=\mathbb C$.

By  $b^*_x$ and $b_x$ we will denote the bosonic creation and annihilation operators, respectively, that satisfy the canonical commutation relations 
\be
\relax [b_x, b_y] = 0, \quad [b^*_x, b^*_y] = 0, \quad [b_x, b^*_y] = \delta_{x,y}.
\ee
The local number operator is defined as $n_x = b^*_x b_x$, and for any subset $X \subset \mathcal V$, the total number operator $N_X$ is given by
\be
N_X = \sum_{x \in X} n_x.
\ee
 Let $\mathcal{B}$ be the algebra of bounded operators on the Fock space and $\mathcal{A}_X \subset \mathcal{B}$ be the algebra of bounded operators supported on $X \subset \mathcal V$. Given an observable $A$, we will refer to the smallest $X$ such that $A\in \mathcal{A}_X$ as the support of $X$.

 \subsection{Hamiltonian and dynamics}
\label{sec:HB}
Let $D\in3\N$. We consider finite-range Bose--Hubbard type Hamiltonians with range parameter $D$. For $X\subset \mathcal V$, the time-dependent  Bose--Hubbard type Hamiltonian on $X$ is given by
\be\label{Hamil-BH}
 H^{\mathrm{BH}}_X(t)=I_X(t)+V_X(t)=\sum_{Z\subset X}\left(I_Z(t)+V_Z(t)\right),
\ee
where  $I_Z(t)$ is of the form
\be
I_Z(t)= \sum_{\substack{x,y\in Z:\\ Z=\{x,y\}}} J_{\langle x,y\rangle}(t)b^*_x b_y+h.c.
\ee
and

\begin{itemize}
    \item $J_{\langle x,y\rangle}(t)$ are complex-valued continuous functions of $t$  that satisfy $J_{\langle x,y\rangle}=0$ for $ d(x,y)> D/3$, so the {\em hopping range} is at most $D/3$.
\item $V_Z(t)$ is a real-valued, locally integrable, time-dependent function of $\set{n_x:x\in Z}$, where only finite range ($=\frac D3$) interactions between number operators are allowed.
\end{itemize}

\begin{remark}
    The rationale behind the choice of $D/3$ here is that after going to the interaction picture, we will produce a Hamiltonian of the form
\be\label{Hamil}
H_X(t)=\sum_{x,y\in X}\mathcal{U}_X(t) J_{\langle x,y\rangle}(t)b^*_x b_y\mathcal{U}_X^*(t)+h.c.
\ee
of interaction range at most $D$. This is  because in the interaction picture the hopping range $D/3$ is spread out over a ball of radius $D/3$; see Section \ref{ssect:interaction_picture}.
\end{remark}

Throughout the paper, we will often suppress the set dependence of quantities when the set in question is $X=\mathcal V$. E.g., we will simply use $H$ instead of  $\tilde H_{\mathcal V}$ and $N$ instead of $N_\mathcal V$ in the sequel.

 We assume that 
\be\label{eq:J}
J:=\sup_t\sup_{x\in\mathcal V}\sum_{y\in\mathcal V}\abs{J_{\langle x,y\rangle}(t)}<\infty.
\ee
In fact, without loss of generality, we will set  $J=1$ by rescaling the Hamiltonian if necessary.

We note that $H^{\mathrm{BH}}(t)$ commutes with  the total particle number operator $N_G=\sum_{x\in G} n_x$ for  all $t$.  Hence, we can represent $H^{\mathrm{BH}}(t)$ as a direct sum of bounded Hamiltonians of systems with a fixed number $N$ of bosons; see, e.g., \cite[App.~A]{FLS2022lieb}. This feature allows us to associate the unitary propagator $U(t,s)$ with the family $\tilde  H(t)$ (the Schr\"odinger  time evolution), given by the initial value problem
 \be\label{Schr}
 i\partial_{t}U(t,s)=   H^{\mathrm{BH}}(t)U(t,s),\quad U(s,s)=\mathds{1}.
 \ee
 $U$ is a function of parameters $s$ and $t$, where $s$ is the start time and $t$ is the stopping time. 
 In general, the existence of dynamics for unbounded operators $H^{\mathrm{BH}}(t)$ requires  argumentation, but in our case $U$ is a direct sum of propagators associated with bounded Hamiltonians, so this does not present a problem.

For the time-independent Hamiltonian $H^{\mathrm{BH}}$ the propagator is given by $U(t,s)=e^{-i(t-s)\tilde  H}$. Some  basic properties of $U(t,s)$ that will be used in this work can be found in Appendix \ref{app:B} below. 
 
 \subsection{Notation}
Given an operator $F$, let $\mathcal{F} \cdot := -i[F, \cdot]$ be the adjoint action of $F$. Since it acts on operators in the Hilber space, it is often called a superoperator.

The Heisenberg time evolution $\tau^{H^{\mathrm{BH}}} A$  of a bounded observable $A$ induced by the time-dependent Hamiltonian $H^{\mathrm{BH}}(t)$ is another superoperator given by $\tau_{t,s}^{H^{\mathrm{BH}}} A=U(t,s)AU^*(t,s)$.  It is a solution of the initial value problem  
 \be\label{Heis}
 \partial_{t}\tau_{t,s}^{H^{\mathrm{BH}}} A =\mathcal{H^{\mathrm{BH}}}(t)\tau_{t,s}^{H^{\mathrm{BH}}} A,\quad \tau_{s,s}^{H^{\mathrm{BH}}} A =A.
 \ee
In the case of a time-independent Hamiltonian $H^{\mathrm{BH}}$, the expression simplifies to
\be
\tau_{t,s}^{H^{\mathrm{BH}}} A= e^{-i(t-s)H^{\mathrm{BH}}} A e^{i(t-s)H^{\mathrm{BH}}}.
\ee

We will let $\circ$ represent the superoperator concatenation over a time interval, 
\be\label{compos}
 C= A\circ  B,\quad  C_{t,s}=( A\circ  B)_{t,s}=\int _{s}^{t} A_{t,r} B_{r,s}\,dr.
\ee

Throughout the paper we will always impose the right to left order of operations for superoperators: In the expression $\mathcal S\mathcal TA$ where $\mathcal S,\mathcal T$ are superoperators and $A$ is an observable,  $\mathcal T$ acts on $A$ first, and $\mathcal{S}$ acts second.

We also define the $R$-enlargement of a subset $Y$ within $X$ (where $Y \subset X \subset \mathcal V$) for any $R > 0$ as
\be
\relax [Y]_R(X)=\{x \in X : d(x, Y) \leq R\}.
\ee
Here, $d(x, Y)$ denotes the graph distance between the vertex $x$ and $Y$. When referring to the enlargement relative to the entire set $\mathcal V$, we use the abbreviated notation $[Y]_R$.

For $ q\geq 0$ and $X\subset\mathcal V$, let $\Pi_{ q }(X)$ be the projection onto
the eigenspace of the boson number operators $\set{n_x}_{x\in X}$ with eigenvalues
smaller than or equal to $ q $, 
\be\label{piq}
\Pi_{ q }(X)=\prod_{x\in X}\chi_{n_x\le  q }\ .
\ee
We will denote by  $|X|$ the cardinality of a set $X$. For $r\in\R$, we will use the Japanese brackets $\langle r \rangle$ for $|r|+1$, and for $x\in\mathcal V$, we will denote  $\langle x\rangle=d(x,0)+1$.

\subsection{Main results}
For $r\in\N$ with $L/(r+2)\in\N$, let
\be\label{eq:r'}
\ell=\lceil\ln r\rceil,\quad Q=\lceil Mqr^d\ell^2\rceil.
\ee
One should think of $\ell$ as unimportant logarithmic corrections, of $M$ and $q$ as order-$1$ parameters, and of $r$ like time $t$.

\begin{theorem}[Main result]\label{kthm}
 Let $H^{\mathrm{BH}}$ be as in \eqref{Hamil-BH} with $J=1$ in \eqref{eq:J}. Then there exist constants $C,c,M_0>0$ depending only on the dimension $d$ and the range $D$ such that for  any bounded observable $A_Y$  and supported on a finite $Y\subset \mathcal V$ with $\|A_Y\|=1$, any $ q \in \N$, and any $M\ge M_0$, we have
\be\label{eq:mbn}
&\norm{\Pi_{ q }( \mathcal V)\pa{ \tau_{t,0}^{H^{\mathrm{BH}}}A_Y-A_Y(t)}\Pi_{ q }( \mathcal V)}\le C\abs{Y}\langle L \rangle^d\langle t \rangle^2Q^2r^{d}\pa{\pa{\frac e\ell}^r+e^{-cr^d}},
\ee
where the observable $A_Y(t)$ is supported on $[Y]_{D(L+1)}$ with $r\ge C\langle t \rangle$ and 
\be\label{eq:L}
L\ge CMq\langle t\rangle r^{d-1}\ell^8.
\ee
\end{theorem}

\begin{remark}[Light cone shape]
We note that the lower bound $r\ge C\langle t \rangle$ implies that the value for $L$ must be at least of order $\langle t\rangle^d$ (up to logarithmic corrections in $t$) to keep the left-hand side in \eqref{eq:mbn} small. This matches the main result $v\sim \langle t \rangle^{d-1}$ in \cite{Kuwahara2024} and is known to be sharp \cite{Kuwahara2024} for the general class of time-dependent Bose-Hubbard Hamiltonians considered here.
\end{remark}

As a consequence, we obtain the following analog of the Kuwahara, Vu, and Saito result, \cite{Kuwahara2024}: Consider an initial state (density matrix) $\rho$ that satisfies $\rho\ge0$, $\tr\rho=1$ and the assumption of bounded  local density,
\be\label{eq:locde}
\sup_{x\in\mathcal V} \tr \pa{\rho e^{(n_x)^\kappa}}<\infty \mbox{ for some } \kappa>0. 
\ee
\begin{corollary}[Application to controlled-density initial states]\label{thm}
 Let $H^{\mathrm{BH}}$ be as in \eqref{Hamil-BH} with $J=1$ in \eqref{eq:J}.
Let $\kappa>0$ and $p>0$. There exists $C_p>0$, depending on $p,d,D,\kappa$ and the bound in \eqref{eq:locde}, such that for any bounded observable $A_Y$  supported on $Y\subset\mathcal V$ with $\|A_Y\|=1$, there exists an observable $A_Y(t)$ supported on $[Y]_{D\langle L \rangle}$, where
\be\label{eq:r''}
L=C_p|t|^d(\ln|t|)^{8+2/\kappa},
\ee
\be
\norm{ \rho^{1/2} \pa{\tau_{t,0}^{H}A_Y- A_Y(t)}\rho^{1/2}}_1 \le C_p\norm{A_Y}\abs{Y}|t|^{-p},
\ee
holds for $|t|\geq 2$ and any  state $\rho$ satisfying \eqref{eq:locde}. 
\end{corollary}

We emphasize that we only require  $A$ to be a bounded local observable. Prior works \cite{Kuwahara2024,kuwahara2026symmetry} additionally assumed that $A$ changes the local particle number by a finite amount which then entered the LRB explicitly. We can avoid this additional assumption by considering the symmetrized norm $\norm{ \rho^{1/2} \pa{\tau_{t,0}^{H^{\mathrm{BH}}}A_Y- A_Y(t)}\rho^{1/2}}_1$ instead of $\norm{ \pa{\tau_{t,0}^{H}A_Y- A_Y(t)}\rho^{1}}_1$.

\section{Bounds on particle transport}\label{sec:conc}
A structural feature of the Bose-Hubbard Hamiltonian $H^{\mathrm{BH}}$ is that its commutator with a number operator $n_x$ is quadratic in creation / annihilation operators. Consequently, the higher order commutators with number operators retain this feature as well, allowing us to obtain a number of estimates pertaining to particle (boson) transport in the course of evolution. The bounds we derive in this section play a role similar to that of the so-called energy estimates in the theory of Schr\"odinger operators. 

\subsection{Interaction picture}
\label{ssect:interaction_picture}
Let $X\subset \mathcal V$. Consider the interaction picture, that is,  a time-dependent Hamiltonian 
\[H_X(t)=\mathcal{U}_X(t)I_X(t) \mathcal{U}_X^*(t)\]
where $\mathcal{U}_X(t)$ is generated by $V_X(t)$,
\be
\partial_t \mathcal{U}_X(t)=iV_X(t)\mathcal{U}_X(t), \quad \mathcal{U}_X(0)=\mathds{1}.
\ee
 $\mathcal{U}_X(t)$ is a unitary operator with range $D/3$ satisfying
 \be
 \relax [n_x,\mathcal{U}_X(t)]=0,\quad \forall x\in \mathcal V.
 \ee
 $H_X(t)$ is again self-adjoint and has an associated Heisenberg dynamics defined via \eqref{Heis} with $\tilde H_X$ replaced by $H_X$.
This Heisenberg dynamics $\tau_{t,0}^{\tilde H_X} A_Y$  
  is related to $\tau_{t,0}^{H_X}  A_Y$  by
\[\tau_{t,0}^{\tilde H_X}A_Y=\mathcal{U}_X^*(t)\pa{\tau_{t,0}^{H_X}A_Y}\mathcal{U}_X(t),\]
see  \cite[Section X.12]{RS} for details. 
The transformed Hamiltonian  $H_X(t)$ is exactly of the form \eqref{Hamil}, i.e.,
\[
H_X(t)=\sum_{x,y\in X}\mathcal{U}_X(t) J_{\langle x,y\rangle}(t)b^*_x b_y\mathcal{U}_X^*(t)+h.c.
\]
Each term $\mathcal{U}_X(t) J_{\langle x,y\rangle}(t)b^*_x b_y\mathcal{U}_X^*(t)$ has range bounded by $3\times \frac{D}{3}=D$.

We will henceforth work with the dynamics generated by \eqref{Hamil}.  In particular, we will prove Theorem \ref{kthm} and Corollary \ref{thm} for these, except that $A_Y(t)$ will be supported on $[Y]_{DL}$. Since $\mathcal{U}_X(t)\Pi_ q (\mathcal V) \mathcal{U}_X^*(t)=\Pi_ q (\mathcal V)$ for any $X\subset \mathcal V$ and unitary transformations preserve the norm,  Theorem \ref{kthm} and Corollary \ref{thm} then immediately follow for  $\tau^{\tilde H_X} $ as well.

We will use some basic properties of Schr\"odinger and Heisenberg dynamics that can be found in Appendix \ref{app:B}.

\subsection{Bounds on particle transport}
For a bounded operator $T$ on $\ell_2(G)$ and $x,y\in\mathcal V$, we will denote by $T(x,y)$ its kernel, 
\be
T(x,y)=\langle \delta_x,T\delta_y\rangle,
\ee
 where $\delta_x$ is the Kronecker delta function, i.e.,  the indicator of site $x$.

Let $T$ be a bounded self-adjoint operator on $\ell_2(G)$. We can associate with $T$ a quadratic form $\hat T$ acting in the Fock space as
\be\label{hatT}
\hat T=\sum_{x,y\in\mathcal V} T(x,y)b^*_yb_x.
\ee
We note that $\hat T$ is self-adjoint, that  if $T$ is diagonal, then $\hat T=\sum_{x\in G}T(x,x)n_x$, and that $\hat T\ge 0$ if $T\ge0$. 

Writing $H=\sum_{Z\subset \mathcal V} h_Z(t)$, we will denote by $H'$ an operator of the form 
\be\label{H'}
H'=\sum'_{Z\subset \mathcal V} h_Z(t),
\ee 
where the sum $\sum'$ runs over some subset of all possible sets $Z$.

The main result of this section is as follows. 

\begin{proposition}[Bound on particle transport]\label{nrm1}
Let $H'$ be of the form \eqref{H'}, and fix $q,R,V\in\N$ with $q\ge1$, $R\ge eD$, and $Z\subset \mathcal V$. Then for any $l,V\in\N$ that satisfies $4eD|t-r|\le R$, $l\le4V$, and $\sum_{x\in\mathcal V}e^{-d(x,Z)/R}\le V$, there exists a constant $C$ such that  
\be\label{smb1}
\norm{\Pi_{ q }( \mathcal V)\tau_{t,r}^{H'} \pa{\hat T_{Z}^R+l}^{V}}\le \pa{CqV}^{V}.
\ee
\end{proposition}

Notice that \eqref{smb1} immediately implies the same bound on the original Bose-Hubbard Hamiltonian:
\[
\norm{\Pi_{ q }( \mathcal V)\tau_{t,r}^{H^{\mathrm{BH}}} \pa{\hat T_{Z}^R+l}^{V}}\le \pa{CqV}^{V}.
\]
The following subsection will prove Proposition \ref{nrm1}. Afterwards, we explain how it can be used to compare the full dynamics to a truncated one.

\subsection{Proof of particle transport bound}

 \begin{lemma}[Commutator inequality]\label{lem:T} 
 Let $R\ge D$,  $Z\subset\mathcal V$,  and $a\ge \frac {eD} R $.  Define $T_{Z}^R$ as a self-adjoint multiplication  operator on $\ell_2(G)$ with
 \be
T_{Z}^R(x,x)=e^{-d(x,Z)/R}. 
 \ee
 Then for any odd $k\in\N$, any choice of skipped sets in $H'$, and any $Z\subset\mathcal V$ the following operator inequality holds:
  \be\label{comopRk}
 i\left[ H'(t), \pa{\hat T_{Z}^R+a}^k\right]\le   \frac{2eDk}R\pa{\hat T_{Z}^R+a'}^{k},\quad a'=a+\frac{eD}R.
 \ee

 \end{lemma}

\begin{proof}[Proof of Lemma \ref{comopRk}]
The proof will use induction in $k$. We will suppress the dependence on $t$ in the notation. We have
\be
i[H', n_x]=i\mathcal{U}\pa{\sum_{u,v\in \mathcal V}'J_{\langle u,v\rangle}\left[b^*_u b_v,n_x\right]}\mathcal{U}^*\ + \ h.c.=i \mathcal{U}\pa{\sum_{u,v\in \mathcal V}'J_{\langle u,v\rangle}(\delta_{xv}-\delta_{xu}) b^*_u b_v} \mathcal{U}^*\ + \ h.c.,
\ee
where in the first step we used $[\mathcal{U},n_x]=0$. 
Consequently, 
\be\label{hatK}
 i[H', \hat T_{Z}^R+a]&=\sum_{x,y\in \mathcal V}'  \mathcal{U}K(x,y) b^*_x b_y\mathcal{U}^*\ + \ h.c., 
\ee
where $K(x,y)$  is given by 
\be\label{eqK}
K(x,y)=-i\pa{e^{-d(x,Z)/R}-e^{-d(y,Z)/R}}J_{\langle x,y\rangle}.
\ee
We will use

\begin{lemma}\label{kerest}
Let $\diam(Z)=\max_{u,v\in Z}d(u,v)$. Then for $d(x,y)\le D$, we have
\be
\abs{e^{-d(x,Z)/R}-e^{-d(y,Z)/R}}\le\frac{eD}Re^{-d(x,Z)/R},
\ee
provided $R\ge D$. 
\end{lemma}
We postpone its proof for now. Applying Lemma \ref{kerest} yields 
\be\label{eqKxy} 
\sum_{y\in \mathcal V}' \abs{K(x,y)}&\le \frac{eD}Re^{-d(x,Z)/R},\quad 
\sum_{x\in \mathcal V}'  \abs{K(x,y)}\le\frac{eD}Re^{-d(y,Z)/R}.
\ee 
Using  the operator inequality $zb^*_x b_y+h.c.\le |z|\pa{n_x+n_y}$ in \eqref{hatK}, we obtain the bound
\be\label{hatK1}
 i[H', \hat T_{Z}^R+a]&\le \sum_{x,y\in \mathcal V}'\abs{K(x,y)}\mathcal{U}\pa{ n_x +n_y} \mathcal{U}^*= \sum_{x,y\in \mathcal V}'  \abs{K(x,y)}\pa{ n_x +n_y} \\&\le\frac{2eD}R\hat T_{Z}^R\le\frac{2eD}R\pa{\hat T_{Z}^R+a'}. 
\ee
This establishes the induction base.

 Suppose that \eqref{comopRk} holds for some $k=2m-1$ with $m\ge1$ (the induction hypothesis), we want to show that it also holds for $k=2m+1$ (the induction step). We have 
 \be\label{deco-commutator}
 &i\left[H', \pa{\hat T_{Z}^R+a}^k\right]=i\left[H', \pa{\hat T_{Z}^R+a}\pa{\hat T_{Z}^R+a}^{2m-1}\pa{\hat T_{Z}^R+a}\right]\\&\hspace{1cm}=i\left[H',\hat T_{Z}^R\right]\pa{\hat T_{Z}^R+a}^{2m}+h.c.+i\pa{\hat T_{Z}^R+a}\left[H', \pa{\hat T_{Z}^R+a}^{2m-1}\right]\pa{\hat T_{Z}^R+a}.
 \ee
 The last contribution on the right hand side we can bound as
   \be\label{ind}
i\pa{\hat T_{Z}^R+a}\left[H', \pa{\hat T_{Z}^R+a}^{2m-1}\right]\pa{\hat T_{Z}^R+a}  \le  \frac{2eD(2m-1)}{R}\pa{\hat T_{Z}^R+a'}^{2m+1}
 \ee
using the induction hypothesis. On the other hand, $i\left[H',\hat T_{Z}^R+a\right]$ is given by \eqref{hatK}. We note that for any polynomial $p$, we have $b_xp(n_x)=p(n_x+1)b_x$, which implies that
\be
\mathcal{U}\pa{b^*_yb_x}\mathcal{U}^*\pa{\hat T_{Z}^R+a}^{2m}&=\mathcal{U}b^*_y\pa{\hat T_{Z}^R+a+T_{Z}^R(x,x)}^{m}b_x\pa{\hat  T_{Z}^R+a}^{m}\mathcal{U}^*\\&=\mathcal{U}\pa{\hat T_{Z}^R+a+T_{Z}^R(x,x)-T_{Z}^R(y,y)}^{m}b^*_yb_x\pa{\hat T_{Z}^R+a}^{m}\mathcal{U}^*,
\ee
using $[\mathcal{U},\hat T_{Z}^R]=0$.

We next observe that using \eqref{eqK1},  for $a\ge \frac {eD} R $, 
\be
\hat T_{Z}^R+a+T_{Z}^R(x,x)-T_{Z}^R(y,y)\ge\hat T_{Z}^R\ge 0.
\ee
Using the operator inequality $A^*B+B^*A\le A^*A+B^*B$ with
\[A=b_y\pa{\hat T_{Z}^R+a+T_{Z}^R(x,x)-T_{Z}^R(y,y)}^{m},\quad B=b_x\pa{\hat T_{Z}^R+a}^{m},\]
we get the bound
\be
\mathcal{U}\pa{b^*_yb_x}\mathcal{U}^*\pa{\hat T_{Z}^R+a}^{2m}+h.c.&\le\pa{\hat T_{Z}^R+a+T_{Z}^R(x,x)-T_{Z}^R(y,y)}^{2m}n_y+\pa{\hat T_{Z}^R+a}^{2m}n_x\\&\le \pa{\hat T_{Z}^R+a'}^{2m}\pa{n_x+n_y},
\ee
where in the first step we also used $[\mathcal{U},n_x]=0$.

In light of  \eqref{hatK} and  \eqref{eqKxy}, we obtain the operator inequality
\be\label{inds}
i\left[H',\hat T_{Z}^R\right]\pa{\hat T_{Z}^R+a}^{2m}+h.c.&\le \frac{4eD}{R}\pa{\hat T_{Z}^R+a'}^{2m}\hat T_{Z}^R\le \frac{4eD}{R}\pa{\hat T_{Z}^R+a'}^{2m+1}
.
\ee
Combining \eqref{deco-commutator}, \eqref{ind}, and \eqref{inds}, we get the induction step.
\end{proof}
\begin{proof}[Proof of Lemma \ref{kerest}] 
We note that the inequality $|e^s - 1| \le e|s|$ holds for all $s \in [-1, 1]$. It implies that for $d(x,y)\le D$, we can bound 
\be\label{eqK1}
\abs{e^{-d(x,Z)/R}-e^{-d(y,Z)/R}}\le\frac{eD}Re^{-d(x,Z)/R},
\ee
since $\abs{d(x,Z)-d(y,Z)}\le d(x,y)$ by the triangle inequality.
\end{proof}
Let $H'$ be of the form \eqref{H'}, and let $\tau_{t,r}^{H'}$ denote the corresponding Heisenberg evolution. 
\begin{corollary}\label{cor} Let $R\ge eD$, then for any $l\in\N$, any $t$, any $Z\subset \mathcal V$, and any observable $A$  the following operator inequalities hold.
\be\label{comexpz1}
\tau_{t,r}^{H'} \pa{\hat T_{Z}^R+l}^k\le  \exp\pa{\frac {4eD|t-r|k}R}\pa{\hat T_{Z}^R+l+\frac{eDk}R}^{k},\quad  k\in 2\N+1;
\ee
\be\label{comexpz2}
\norm{\pa{\hat T_{Z}^R+l+\frac{eD(2k+1)}R}^{-k}\tau_{t,r}^{H'}\pa{\pa{\hat T_{Z}^R+l}^kA}}\le  \exp\pa{\frac {4eD|t-r|k}R}\norm{A},\quad  k\in \N.
\ee
\end{corollary}
\begin{proof}
We first observe that \eqref{comexpz2} is a direct consequence of  \eqref{comexpz1}:
\be
&\norm{\pa{\hat T_{Z}^R+l+\frac{eD(2k+1)}R}^{-k}\tau_{t,r}^{H'}\pa{\pa{\hat T_{Z}^R+l}^kA}}^2
\\&=\norm{\pa{\hat T_{Z}^R+l+\frac{eD(2k+1)}R}^{-k}
\tau_{t,r}^{H'}\pa{\pa{\hat T_{Z}^R+l}^k 
AA^* \pa{\hat T_{Z}^R+l}^{k}}\pa{\hat T_{Z}^R+l+\frac{eD(2k+1)}R}^{-k}}
\\&\le\norm{A}^2\norm{\pa{\hat T_{Z}^R+l+\frac{eD(2k+1)}R}^{-k}\pa{\tau_{t,r}^{H'} \pa{\hat T_{Z}^R+l}^{2k}}
\pa{\hat T_{Z}^R+l+\frac{eD(2k+1)}R}^{-k}}\\&\le  \exp\pa{\frac {8eD|t-r|k}R}\norm{A}^2,
\ee
where in the last step we used \eqref{comexpz1} with $k$ replaced by $2k+1$ and the operator-monotonicity of $x\mapsto x^{2k/(2k+1)}$.
%

On the other hand, to show \eqref{comexpz1} It suffices to verify that it holds for each sector $N_{\mathcal V}=N\in\N$. Consider the case $t>r$, the case $t<r$ can be treated analogously. 

By the fundamental theorem of calculus, for any observable $B$
 \[\tau_{t,r}^{H'} B=B-\int_r^t\partial_s\pa{\tau_{t,s}^{H'} B}ds=B+\int_r^t
\tau_{t,s}^{H'} \mathcal H'(s)B\,ds.\] 
 Using it with $B=\pa{\hat T_{Z}^R+l}^{k}$, $l\in\N$, by the previous lemma applied to $-H'$, we have 
\be
\tau_{t,r}^{H'} \pa{\hat T_{Z}^R+l}^k \le \frac{2eDk}{R} \int_r^t\tau_{t,s}^{H'}\pa{\hat T_{Z}^R+l+\frac{eD}R}^k\,ds+\pa{\hat T_{Z}^R+l}^{k}.
\ee
Iterating, we get that for any $p\in\N$  
\be\label{itera'}
\tau_{t,r}^{H'} \pa{\hat T_{Z}^R+l}^k&\le \sum_{m=0}^{p-1} \frac1{m!}\pa{\frac{2eDk|t-r|}{R}}^m \pa{\hat T_{Z}^R+l+\frac{eDm}R}^{k}\\&\hspace{-2cm}+\pa{\frac{2eDk}{R}}^p\int_{r\le s_p\le\ldots\le s_1\le t}\tau_{t,s_p}^{H'}\pa{\hat T_{Z}^R+l+\frac{eDp}R}^k\,  d\vec s,
\ee
where $\vec s=(s_1,\ldots,s_p)$.

Since  for a sector $N_{\mathcal V}=N$ we have $n_x\le N$ for each $x$, we see that in this sector 
\be
\norm{\pa{\hat T_{Z}^R+l+\frac{eDp}R}^{k}}\le \pa{N+l+\frac{eDp}R}^{k}.
\ee
 On the other hand, $\norm{\tau_{t,s}^{H'} A}=\norm{A}$ for any $A$ and $t,s$, so we can bound the integral term on the right hand side of \eqref{itera'} in the norm by
 \be
\pa{\frac{2eDk|t-r|}{R}}^p\frac1{p!}\pa{N+l+\frac{eDp}R}^{k},
\ee
which converges to zero as $p\to\infty$. This implies 
\be
\tau_{t,r}^{H'} \pa{\hat T_{Z}^R+l}^k\le \sum_{m=0}^\infty  \frac1{m!}\pa{\frac{2eDk|t-r|}{R}}^m \pa{\hat T_{Z}^R+l+\frac{eDm}R}^{k}.
\ee
Since  $2^w>w$ for $w>0$, we see that for any $x\ge1$ and $a,b\ge0$, 
\be
\sum _{m=0}^{\infty }\frac{1}{m!}(ak)^{m}(x+bm)^{k}\le \sum _{m=0}^{\infty }\frac{1}{m!}(2ak)^{m}(x+bk)^{k}= e^{2ak}(x+bk)^{k}.
\ee
Using this with $x=\hat T_{Z}^R+l$, $a=\frac{2eD|t-r|}{R}$, and $b=\frac{eD}R$, we arrive at \eqref{comexpz1}.
\end{proof}

\begin{proof}[Proof of Proposition \ref{nrm1}]
We will suppress the $\mathcal V$  dependence through the proof. We note that
\be
\norm{\Pi_{ q }\tau_{t,r}^{H'}\pa{\hat T_{Z}^R+l}^p}^2=\norm{\Pi_{ q }\pa{\tau_{t,r}^{H'} \pa{\hat T_{Z}^R+l}^{2p}}\Pi_{ q }}.
\ee
Applying the even-moment consequence of Corollary \ref{cor}, the right hand side is bounded by 
\be\label{expa}
 e^{2p} \norm{\Pi_{ q }\pa{\hat T_{Z}^R+2p+l+1}^{2p}\Pi_{ q }},
\ee
using our assumption $4eD|t-r|\le R$.

Let
\be
M_{q,R}=q\sum_{x\in\mathcal V}e^{-d(x,Z)/R}.
\ee
We note that $
M_{q,R}\le qV
$ 
by assumption. 
Since operators $\Pi_{ q }$ and $\hat T_{Z}^R$ are diagonal in the canonical basis, we see from the definition of $\hat T_{Z}^R$ that
\be\label{M_q}
\norm{\Pi_{ q }\pa{\hat T_{Z}^R+2p+l+1}^{2p}\Pi_{ q }}=\pa{M_{q,R}+2p+l+1}^{2p}.
\ee

We now pick $p=V$ so that 
$
M_{q,R}+2p+l+1\le 8qV$. 
With this choice, we get \eqref{smb1}.
\end{proof}

\begin{remark}\label{rema}
If $Z\subset[U]_{C_Dr}$, $\diam U\le D$,  one can use the value $V=\lceil C_{d,D}R^d\rceil$ in \eqref{smb1}. Indeed, for $u\in U$, $d(x,Z)\ge d(x,u)-C_Dr-D$, so the inequality $\sum_xe^{-d(x,u)/R}\le C_dr^d$ yields  
\be
\sum_{x\in\mathcal V}e^{-d(x,Z)/R}\le C_{d,D}R^d\,.
\ee
\end{remark}

\subsection{Comparison to truncated dynamics}
Let $H_1(t)$, $H_2(t)$ be a pair of time-dependent Hamiltonians, with $K(t)= H_1(t)-H_2(t)$. Consider the function
\be
F(s) =  \tau_{t,s}^{H_2}\tau_{s,r}^{H_1}  A
,\quad  r\le s\le t.
\ee
Then $F(r)= \tau_{t,r}^{H_2}  A$,  $F(t)= \tau_{t,r}^{H_1}   A$, and
\be
\partial_s F(s) &=  -\tau_{t,s}^{H_2}\mathcal H_2(s)\tau_{s,r}^{H_1} A +\tau_{t,s}^{H_2}\mathcal H_1(s)\tau_{s,r}^{H_1} A \\& = 
\tau_{t,s}^{H_2}\mathcal K(s)\tau_{s,r}^{H_1}  A .
\ee
By the fundamental theorem of calculus, 
\be\label{duhnn}
\tau_{t,r}^{H_1}   A - \tau_{t,r}^{H_2}  A=  F(t)-F(r)= \pa{\tau^{H_2} \circ \mathcal K \tau^{H_1}}_{t,r}  A,
\ee
where the superoperator concatenation operation $\circ$ was defined in \eqref{compos}.

Let $Q\in\N$  be fixed and for $X\subset \mathcal V$ set 
\be\label{piQ}
 \bar \Pi_{Q}(X)=\mathds{1}-\Pi_{Q}(X),
\ee
where $\Pi_Q(X)$ was introduced in \eqref{piq}.

We define a  truncated operator:
\be\label{eq:tildeH}
\tilde H'(t)=\sum'_{Z\not\subset  [Y]_{D\langle L\rangle}} h_Z(t)+\sum'_{Z\subset  [Y]_{D\langle L\rangle}}\tilde h_Z(t),\quad \tilde h_Z = \Pi_{Q}(Z) h_Z\Pi_{Q}(Z).
\ee

We note that  $\tilde  H(t)$ still commutes with the total number operator $N_\mathcal V$ and  $[\tilde  h_Z(t),\tilde h_{Z'}(t)]=0$ for $Z\cap Z'=\emptyset$. The commutator and moment bounds above also hold for $\tilde H'$ as all boson occupation projections $\Pi_{Q}(Z)$ commute with the $\hat T$ operators, and inserting $\Pi_{Q}(Z)$ do not affect the absolute hopping estimates above. In addition, we have
\be
0\le \bar \Pi_{Q}(X)\le \sum_{x\in X}\chi_{n_x> Q}.
\ee
Let $I_t$ stand for $[0,t]$ if $t\ge0$ and for $[t,0]$ if $t<0$.

In this subsection, we will prove.

\begin{proposition}\label{prop:truncate}
Consider any observable $A_Y$ supported on $Y$. For $q,R\in\N$ as in Lemma \ref{nrm1}, $4eD|t|\le R$, and $V\in\N$ with $V\ge C_{d,D}R^d$, we have \be\label{trun1}
\norm{\Pi_{ q }( \mathcal V)\pa{ \tau_{t,0}^{H'}A-\tau_{t,0}^{\tilde H'} A}\Pi_{ q }( \mathcal V)} \le C_{d,D}\norm{A}|t|  \abs{ [Y]_{D\langle L\rangle}}\pa{CqV}^{V}Q^{1-V}.
\ee
\end{proposition}

\begin{proof}[Proof of Proposition \ref{prop:truncate}]
Choosing $X=[Y]_L$, denoting by $\tilde \tau^{H'} $ the Heisenberg evolution for $\tilde H'(t)$, and using \eqref{duhnn} with  $H_2= H'$, $H_1=\tilde H'$, we get 
\begin{multline}\label{trun}
\norm{\Pi_{ q }( \mathcal V)\pa{ \tau_{t,0}^{H'} A-\tau_{t,0}^{\tilde H'} A}\Pi_{ q }( \mathcal V)}\\ \le 2|t|\norm{A}\max_{s\in I_t}\sum'_{Z\subset [Y]_{D\langle L\rangle}}\norm{\Pi_{ q }( \mathcal V) \tau_{t,s}^{H'}\pa{h_Z(s)-\Pi_{Q}(Z)h_Z(s)\Pi_{Q}(Z)}}.
\end{multline}

%
%
%
We can estimate the right hand side of \eqref{trun} using \eqref{smb1}. Specifically, we have the following result.
\begin{lemma}\label{lemma:truncate}
For $q,R\in\N$ as in Lemma \ref{nrm1}, $4eD|t|\le R$, and $V\in\N$ with $V\ge C_{d,D}R^d$, 
\be\label{trest}
\max_{s\in  I_t}\sum'_{Z\subset [Y]_{D\langle L\rangle}}\norm{\Pi_{ q }( \mathcal V) \tau_{t,s}^{H'}\pa{h_Z(s)-\Pi_{Q}(Z)h_Z(s)\Pi_{Q}(Z)}}\le C_{d,D}\abs{ [Y]_{D\langle L\rangle}}\pa{CqV}^VQ^{1-V}
\ee
and for $\diam(Z)\le D$, 
\be\label{trest1}
\max_{s\in I_t}\norm{\Pi_{ q }( \mathcal V) \tau_{t,s}^{H'}{\bar \Pi_{Q}(Z)}}\le C_{d,D}\pa{CqV}^VQ^{-V}.
\ee
\end{lemma}

Proposition \ref{prop:truncate} follows from this by combining \eqref{trun} and \eqref{trest}. It thus remains to prove the lemma.\end{proof}

\begin{proof}[Proof of Lemma \ref{lemma:truncate}]
We will only establish the first bound, the proof of the second is similar (and easier). 
We can bound
\be
&\norm{\Pi_{ q }( \mathcal V)\tau_{t,s}^{H'}\pa{h_Z-\Pi_{Q}(Z)h_Z\Pi_{Q}(Z)}}\\
&\le
 \norm{\Pi_{ q }( \mathcal V)\tau_{t,s}^{H'}\pa{\hat T_{Z}^R+1}^{V}}
 \norm{\pa{\hat T_{Z}^R+1}^{-V}\pa{h_Z-\Pi_{Q}(Z)h_Z\Pi_{Q}(Z)}}.
\ee
The first factor is bounded by \eqref{smb1}. To estimate the second factor, we use
\be
h_Z-\Pi_{Q}(Z)h_Z\Pi_{Q}(Z)&=\bar \Pi_{Q}(Z)h_Z+\Pi_{Q}(Z)h_Z\bar \Pi_{Q}(Z),\quad [\hat T_{Z}^R,h_Z]=0,\\
\norm{\pa{\hat T_{Z}^R+1}^{-V}\pa{h_Z-\Pi_{Q}(Z)h_Z\Pi_{Q}(Z)}}&\le2\norm{\pa{\hat T_{Z}^R+1}^{-V}\bar \Pi_{Q}(Z)h_Z}.
\ee
For $x,y\in Z$, letting $B=\pa{\hat T_{Z}^R+1}^{-V}\bar \Pi_{Q}(Z)$, we get 
\be\label{Tsq}
\norm{Bb_x^*b_y}^2&=\norm{Bb_x^*b_yb_y^*b_xB}\le\norm{Bn_x(n_y+1)B}=\norm{B^2n_x(n_y+1)}\le\norm{B^2N_Z(N_Z+1)}.
\ee
Since $\hat T_{Z}^R+1\ge N_Z+1$ and 
$
\pa{\hat T_{Z}^R+1}\bar \Pi_{Q}(Z)\ge Q\bar \Pi_{Q}(Z)
$ 
by the definition of $\hat T_{Z}^R$, we deduce that
\be\label{smb2}
\norm{\pa{\hat T_{Z}^R+1}^{-V}\bar \Pi_{Q}(Z)b_x^*b_y}^2&\le\norm{\pa{\hat T_{Z}^R+1}^{2-2V}\bar \Pi_{Q}(Z)}\le Q^{2-2V}.
\ee
Since $[\mathcal U,B]=0$, summing the hopping terms in $h_Z$ gives
\be
\norm{\pa{\hat T_{Z}^R+1}^{-V}\pa{h_Z-\Pi_{Q}(Z)h_Z\Pi_{Q}(Z)}}
&\le2\sum_{x,y\in Z}\abs{J_{\langle x,y\rangle}}\pa{\norm{Bb_x^*b_y}+\norm{Bb_y^*b_x}}\\
&\le4Q^{1-V}\sum_{x,y\in Z}\abs{J_{\langle x,y\rangle}}.
\ee
Combining  \eqref{smb1} and  \eqref{smb2} as well as  summing over $x,y$ and using \eqref{eq:J}, we get \eqref{trest}.
\end{proof}

\subsection{Weighted bounds}
We will also need the following result. 
For $X\subset \mathcal V$,  let 
\be\label{eq:tildeH1}
\tilde H_X(t)=\sum_{x,y\in X}\mathcal{U}(t) J_{\langle x,y\rangle}(t) \Pi_{Q}(\set{x,y})b^*_x b_y\Pi_{Q}(\set{x,y})\mathcal{U}^*(t)+h.c.
\ee

\begin{lemma}\label{lem1a} 
Let $R\ge eD$, then for any $X,W\subset \mathcal V$ and  any $l,k\in\N$ with $l>1$
\be\label{comexpz1b}
\norm{\pa{\hat T_{W}^R+l}^{-k-1} \tilde H_X \pa{\hat T_{W}^R+l}^k}&\le  \pa{1+e^{\frac{eD}R\frac {2k+1} l }},\quad  X\subset W;\\ 
\norm{\pa{\hat T_{W}^R+l}^{-k} \tilde H_X \pa{\hat T_{W}^R+l}^k}&\le 
2Q|X|e^{\frac{eDk}{Rl}}.
\ee
\end{lemma}
\begin{proof}
Let $\delta\in\set{0,1}$.
For $u,v\in X$, we have
\be
&\pa{\hat T_{W}^R+l}^{-k-\delta}\mathcal{U}\pa{b^*_ub_v}\mathcal{U}^*\pa{\hat T_{W}^R+l}^k\\
&\hspace{.3cm}=\mathcal{U}\pa{\hat T_{W}^R+l}^{-\delta/2}b^*_ub_v\pa{\hat T_{W}^R+l+T_{W}^R(u,u)-T_{W}^R(v,v)}^{-k-\delta/2}\pa{\hat T_{W}^R+l}^k\mathcal{U}^*,
\ee
Let
\be
A_{u,v}&=b_u\Pi_{Q}(\set{u,v})\pa{\hat T_{W}^R+l}^{-\delta/2}\mathcal{U}^*,\\
B_{u,v}&=b_v\Pi_{Q}(\set{u,v})\pa{\hat T_{W}^R+l+T_{W}^R(u,u)-T_{W}^R(v,v)}^{-k-\delta/2}\pa{\hat T_{W}^R+l}^k\mathcal{U}^*.
\ee
Using \eqref{eq:tildeH1} and applying the operator Cauchy--Schwarz inequality we can bound
\be
\norm{\sum_{u,v\in X}J_{\langle u,v\rangle}A_{u,v}^*B_{u,v}}
\le \norm{\sum_{u,v\in X}\abs{J_{\langle u,v\rangle}}A_{u,v}^*A_{u,v}}^{1/2}
\norm{\sum_{u,v\in X}\abs{J_{\langle u,v\rangle}}B_{u,v}^*B_{u,v}}^{1/2}.
\ee
Using \eqref{eq:J}, we get
\be\label{eqnu}
\sum_{u,v\in X}\abs{J_{\langle u,v\rangle}}A_{u,v}^*A_{u,v}
&=\sum_{u,v\in X}\abs{J_{\langle u,v\rangle}}\pa{\hat T_{W}^R+l}^{-\delta}\Pi_{Q}(\set{u,v})n_u\Pi_{Q}(\set{u,v})\\
&\le\begin{cases}
\pa{\hat T_{W}^R+l}^{-1}N_X\le\mathds{1},&\delta=1,\ X\subset W,\\
Q\abs{X}\mathds{1},&\delta=0.
\end{cases}
\ee
On the other hand, using $J_{\langle u,v\rangle}=0$ for $d(u,v)>D$ and Lemma~\ref{kerest}, on $\{n_v\ge1\}$ we have
\be
\hat T_{W}^R+l+T_W^R(u,u)-T_W^R(v,v)&\ge l,\\
\frac{\hat T_{W}^R+l}{\hat T_{W}^R+l+T_W^R(u,u)-T_W^R(v,v)}
&\le1+\frac{eD}{Rl}\le e^{\frac{eD}{Rl}}.
\ee
For $X\subset W$, $T_W^R(u,u)=T_W^R(v,v)=1$  since $u,v\in X$. Consequently,
\begin{multline}\label{eqnv}
\sum_{u,v\in X}\abs{J_{\langle u,v\rangle}}B_{u,v}^*B_{u,v}\\
=\sum_{u,v\in X}\abs{J_{\langle u,v\rangle}}\pa{\hat T_{W}^R+l}^{2k}
\pa{\hat T_{W}^R+l+T_W^R(u,u)-T_W^R(v,v)}^{-2k-\delta}\Pi_{Q}(\set{u,v})n_v\Pi_{Q}(\set{u,v})\\
\le\begin{cases}
\pa{\hat T_{W}^R+l}^{-1}N_X\le\mathds{1},&\delta=1,\ X\subset W,\\
e^{\frac{2eDk}{Rl}}\displaystyle\sum_{u,v\in X}\abs{J_{\langle u,v\rangle}}\Pi_{Q}(\set{u,v})n_v\Pi_{Q}(\set{u,v})
\le e^{\frac{2eDk}{Rl}}Q\abs{X}\mathds{1},&\delta=0.
\end{cases}
\end{multline}
Combining \eqref{eqnu} and \eqref{eqnv}, we get
\be
\norm{\pa{\hat T_{W}^R+l}^{-k-\delta}\tilde H_X\pa{\hat T_{W}^R+l}^k}
\le\begin{cases}
2,&\delta=1,\ X\subset W,\\
2Q\abs{X}e^{\frac{eDk}{Rl}},&\delta=0.
\end{cases}
\ee
This proves the claim.
\end{proof}

\section{Reduction to the ball}
The first step in establishing the light cone estimates is a reduction of the problem to  spherical geometry, where the original observable $A_Y$ is replaced by the local interaction term $\tilde h_U$ and the problem is to control the spread of information outside the set $[U]_r$ during time $s$, where $s=t(r+2)/L$ for some strategically  chosen integer parameter $r$. This reduction could be thought of as an analog of the Huygens principle (every point on a wave front acts as a source of secondary spherical wavelets that spread out in the forward direction).

We  will consider $t>0$ here; the $t<0$ case follows by time reversal. For $l\in\N$ satisfying $l\le L$, recalling \eqref{eq:tildeH}, set 
 \be\label{H_L} 
 \tilde H_l(t)=\sum_{Z\subset[Y]_{Dl}} \tilde h_Z(t).
 \ee
We note that for any observable $A$ supported on $[Y]_{Dl}$,  its evolution $\tau^{\tilde H_l} A$ is supported on $[Y]_{Dl}$.  
 
We split $[0,L]$ and $[0,t]$ onto $L/(r+2)$ equal length intervals 
\be
\set{[x_k,x_{k+1}]}_{k=0}^{L/(r+2)-1},\quad \set{[t_k,t_{k+1}]}_{k=0}^{L/(r+2)-1}
\ee 
with $x_0=0$, $x_{L/(r+2)}=L$, $t_0=0$, and $t_{L/(r+2)}=t$ (we will make the suitable choices for parameters $L,r$ later), define $\tilde H_k=\tilde H_{x_k}$ for $k\in[1,L/(r+2)]$, and decompose (with larger indices on the left in the products)
\be\label{decex'}
\tau_{t,0}^{\tilde H}  A_Y&=\pa{\prod_{j=1}^{L/(r+2)}\tau_{t_j,t_{j-1}}^{\tilde H_j} } 
A_Y+\sum_{k=1}^{L/(r+2)} 
\tau_{t,t_k}^{\tilde H}\pa{\tau_{t_k,t_{k-1}}^{\tilde H} -\tau_{t_k,t_{k-1}}^{\tilde H_k} }\pa{\prod_{j=1}^{k-1}\tau_{t_j,t_{j-1}}^{\tilde H_j} } A_Y.
\ee
We note that the operator 
\be
B_k=\pa{\prod_{j=1}^{k-1}\tau_{t_j,t_{j-1}}^{\tilde H_j} } A_Y
\ee
is supported on $[Y]_{Dx_{k-1}}$, so the first term in our decomposition is supported on $[Y]_{DL}$. Using \eqref{duhnn}, we can express
\be
\tau_{t,t_k}^{\tilde H}\pa{\tau_{t_k,t_{k-1}}^{\tilde H} -\tau_{t_k,t_{k-1}}^{\tilde H_k} }B_k&= \tau_{t,t_k}^{\tilde H}\pa{\tau^{\tilde H}\circ\pa{\tilde{\mathcal H}-\tilde{\mathcal H}_k}\tau^{\tilde H_k}}_{t_k,t_{k-1}}B_k
\\ &= \sum_{\substack{U\cap [Y]_{Dx_k}\neq\emptyset\\U\cap [Y]_{Dx_k}^c\neq\emptyset}}
\tau_{t,t_k}^{\tilde H}\pa{\tau^{\tilde H}\circ\tilde{\mathcal{h}}_U\tau^{\tilde H_k}}_{t_k,t_{k-1}}B_k.
\ee
We can rewrite
\be
\tilde{\mathcal{h}}_U(s)\tau^{\tilde H_k}_{s,t_{k-1}}B_k=-\tau^{\tilde H_k}_{s,t_{k-1}}\pa{\mathcal{B}_k \tau^{\tilde H_k}_{t_{k-1},s} \tilde h_U(s)}.
\ee
%
We aim to show that 
\be\label{Hyugens}
\mathcal{B}_k \tau^{\tilde H_k}_{t_{k-1},s} \tilde h_U(s)
\ee
is essentially exponentially small in $r$, which will be enough to control the whole expansion in \eqref{decex'}. Intuitively, it could be small since the support of $B_k$ is at distance greater than $D(r+1)$ from  $U$, so if the time interval $s-t_{k-1}\le t_k-t_{k-1}=t(r+2)/L$ is not too large, then the support of the term $ \tau^{\tilde H_k}_{t_{k-1},s} \tilde h_U$ remains mostly within the distance $Dr$ from $U$ and hence the expression in \eqref{Hyugens} is  small (the summation over all $U$ will result in volume factor). Specifically, for any $s\in[t_{k-1},t_k]$ we can bound
\begin{multline}\label{ges}
\norm{\Pi_q( \mathcal V)\pa{\tau_{t,s}^{\tilde H} 
\pa{\tilde{\mathcal H}(s)-\tilde{\mathcal H}_k(s)}
\tau_{s,t_{k-1}}^{\tilde H_k}  B_k}\Pi_q( \mathcal V)}\\ \le \sum_{\substack{U\cap [Y]_{Dx_k}\neq\emptyset\\U\cap [Y]_{Dx_k}^c\neq\emptyset}}\norm{\Pi_q( \mathcal V) \pa{\tau_{t,s}^{\tilde H}\tau_{s,t_{k-1}}^{\tilde H_k} \pa{\mathcal{B}_k \tau_{t_{k-1},s}^{\tilde H_k} \tilde h_U(s)}}\Pi_q( \mathcal V)}
\end{multline}
for any $s\in[t_{k-1},t_k]$.

Finally, we introduce the $k,U$-dependent Hamiltonian 
\be\label{Wk}
W_k(t)=\sum_{Z\subset[Y]_{Dx_k}\cap [U]_{Dr}} \tilde h_Z(t).
\ee
Using \eqref{duhnn}, we get
\be
\mathcal{B}_k \tau_{t_{k-1},s}^{\tilde H_k}  \tilde h_U&=\mathcal{B}_k \tau_{t_{k-1},s}^{W_k} \tilde h_U+\mathcal{B}_k\pa{ \tau^{\tilde H_k} \circ \pa{\tilde{\mathcal H}_k-\mathcal W_k}\tau^{W_k}}_{t_{k-1},s} \tilde h_U
\\&=\sum_{\substack{Z_1\subset[Y]_{Dx_k}\\Z_1\not \subset [U]_{Dr},\ Z_1\not\subset [U]_{Dr}^c}} \mathcal{B}_k\pa{ \tau^{\tilde H_k} \circ \tilde{\mathcal{h}}_{Z_1}\tau^{W_k}}_{t_{k-1},s} \tilde h_U,
\ee
where in the second step, we used the fact that $\tau^{W_k} \tilde h_U$ is supported outside of $\supp(B_k)$. 

Denoting
\be\label{Dk}
D_k(u,s)=\tilde{\mathcal{h}}_{Z_1}(u) \tau_{u,s}^{W_k} \tilde h_U(s),
\ee
this leads to the bound
\begin{multline}\label{ges1}
\norm{\Pi_q( \mathcal V) \pa{\tau_{t,s}^{\tilde H}\tau_{s,t_{k-1}}^{\tilde H_k} \pa{\mathcal{B}_k \tau_{t_{k-1},s}^{\tilde H_k}\tilde h_U(s)}}\Pi_q( \mathcal V)}
\\ \le \sum_{\substack{Z_1\subset[Y]_{Dx_k}\\Z_1\not \subset [U]_{Dr},\ Z_1\not\subset [U]_{Dr}^c}} \norm{\Pi_q( \mathcal V) \pa{\tau_{t,s}^{\tilde H}\tau_{s,t_{k-1}}^{\tilde H_k}\mathcal{B}_k \pa{\tau^{\tilde H_k}\circ D_k}_{t_{k-1},s} }\Pi_q( \mathcal V)}.
\end{multline}
In the following argument, we will use the constraint  
\be\label{constr}
r>4eD |t|.
\ee
Let $\Theta=[U]_{D(r+1)}$,  $\xi=qr^{d}M$, and $V=\lceil C_{d,D}r^d\rceil$, with $C_{d,D}$ chosen so that $\sum_xe^{-d(x,\Theta)/r}\le V$. The operator $D_k(u,s)$ is supported on~$\Theta$, commutes with~$N_\Theta$, and satisfies $\norm{D_k(u,s)}\le CQ^2$. We can bound
\be\label{redb1}
\norm{\Pi_q\tau^{\tilde H}_{t,s}\tau^{\tilde H_k}_{s,t_{k-1}}\pa{\mathcal B_k\tau^{\tilde H_k}_{t_{k-1},u}D_k(u,s)}\Pi_q}
&\le\norm{\Pi_q\tau^{\tilde H}_{t,s}\tau^{\tilde H_k}_{s,t_{k-1}}\pa{\mathcal B_k\tau^{\tilde H_k}_{t_{k-1},u}\pa{D_k(u,s)\chi_{N_\Theta<\xi}}}\Pi_q}\\
&+\norm{\Pi_q\tau^{\tilde H}_{t,s}\tau^{\tilde H_k}_{s,t_{k-1}}\pa{\mathcal B_k\tau^{\tilde H_k}_{t_{k-1},u}\pa{D_k(u,s)\chi_{N_\Theta\ge\xi}}}\Pi_q}.
\ee
To estimate the second term, we recall that  $\tau^{H} A=U(t,s)AU^*(t,s)$, where $U(t,s)$ satisfies \eqref{Schr}. Denoting by $\tilde U$ and $\tilde U_k$ the corresponding evolutions generated by $\tilde H$ and $\tilde H_k$ respectively, we can bound
\be
\norm{\chi_{N_\Theta\ge\xi}\tilde U_k(u,s)\tilde U(s,t)\Pi_q}
&\le\norm{\chi_{N_\Theta\ge\xi}\pa{\hat T_\Theta^r+1}^{-V}}
\norm{\pa{\hat T_\Theta^r+1}^{V}\tilde U_k(u,s)\pa{\hat T_\Theta^r+2V+2}^{-V}}\\
&\quad\times\norm{\pa{\hat T_\Theta^r+2V+2}^{V}\tilde U(s,t)\Pi_q}.
\ee
Using
\be
\norm{\chi_{N_\Theta\ge\xi}\pa{\hat T_\Theta^r+1}^{-V}}\le\xi^{-V},
\ee
\eqref{comexpz2} for the adjoint propagator, and \eqref{smb1}, we obtain the estimate
\be
\norm{\chi_{N_\Theta\ge\xi}\tilde U_k(u,s)\tilde U(s,t)\Pi_q}
&\le\xi^{-V}e^{4eDV|s-u|/r}\pa{CqV}^{V}\le e^{-V}
\ee
for $M$ large enough (but $r$ and $q$ independent). Using the fact that $N_{\Theta}$ commutes with $D_k(u,s)$, expanding the commutator and using
$
\tau^{\tilde H}_{t,s}\tau^{\tilde H_k}_{s,t_{k-1}}\tau^{\tilde H_k}_{t_{k-1},u}
=\tau^{\tilde H}_{t,s}\tau^{\tilde H_k}_{s,u}
$
gives
\be
\norm{\Pi_q\tau^{\tilde H}_{t,s}\tau^{\tilde H_k}_{s,t_{k-1}}\pa{\mathcal B_k\tau^{\tilde H_k}_{t_{k-1},u}\pa{D_k(u,s)\chi_{N_\Theta\ge\xi}}}\Pi_q}
&\le2\norm{B_k}\norm{D_k(u,s)}\norm{\chi_{N_\Theta\ge\xi}\tilde U_k(u,s)\tilde U(s,t)\Pi_q}\\
&\le C\norm{A_Y}Q^2e^{-V}.
\ee
For the first term in~\eqref{redb1},
\be
\norm{\Pi_q\tau^{\tilde H}_{t,s}\tau^{\tilde H_k}_{s,t_{k-1}}\pa{\mathcal B_k\tau^{\tilde H_k}_{t_{k-1},u}\pa{D_k(u,s)\chi_{N_\Theta<\xi}}}\Pi_q}
\le2\norm{A_Y}\norm{D_k(u,s)\chi_{N_\Theta<\xi}}.
\ee
Plugging these into \eqref{redb1}, we see that 
\be\label{redb2}
\norm{\Pi_q\tau^{\tilde H}_{t,s}\tau^{\tilde H_k}_{s,t_{k-1}}
\pa{\mathcal B_k\tau^{\tilde H_k}_{t_{k-1},u}D_k(u,s)}\Pi_q}
\le C\norm{A_Y}\pa{\norm{D_k(u,s)\chi_{N_\Theta<\xi}}+Q^2e^{-V}}\,.
\ee
We reduced the problem to the norm estimation of $D_k(u,s)\chi_{N_{\Theta}<  \xi}$, which is supported on the ball $[U]_{D(r+1)}$. 

\section{Duhamel expansion}
Although it is possible to derive the light cone $L\sim|t|^{d+1}$ using boson concentration estimates obtained in Section \ref{sec:conc} together with the standard Lieb--Robinson approach for spin systems, establishing the sharp  bound $L\sim|t|^d$ requires a refinement of the  latter theory. The first step toward this goal is the development of a Duhamel path expansion for the object of interest (specifically, the quantity $D_k$ defined in \eqref{Dk}). We then will use it to rewrite $D_k$ in a form amenable to further analysis by introducing a number of boson number cutoffs along the path.

\subsection{The radial expansion}
 Let   $r\in\N$ and let subsets $U,Z_1\subset\mathcal V$ with $\diam U\le D$ and $\diam Z_1\le D$ be such that $Z_1\not \subset [U]_{Dr}$ and $Z_1\not\subset [U]_{Dr}^c$, i.e., $Z_1$ lies at the boundary of $[U]_{Dr}$. Let $S$ be a set that is contained in $[U]_{Dr}$ and intersects $Z_1$. Recall that $\tilde h_Z(t) = \Pi_{Q}(Z) h_Z(t)\Pi_{Q}(Z)$.

 For $j$ satisfying $j\le r$, we set 
 \be\label{H_j} 
 \tilde H_j(t)=\sum_{Z\subset[U]_{Dj}\cap S} \tilde h_Z(t),\quad K_j(t)= -i\pa{\tilde H_j(t)- \tilde H_{j-1}(t)},\qquad \mathcal K_j A=[K_j,A].
 \ee
We note that for any observable $A$ supported on $[U]_{Dl}$,  its evolution $\tau^{\tilde H_l} A$ is supported on $[U]_{Dl}$.  We will use the shorthand notation $\tau^j$ for $\tau^{\tilde H_j}$ in this section to simplify notation. Furthermore, we will suppress the time dependence to make the resulting expressions  compact. 

In this notation, \eqref{duhnn} yields
\be
\tilde{\mathcal{h}}_{Z_1} \tau^{r} \tilde h_U
 &=\tilde{\mathcal{h}}_{Z_1} \left[ \tau^{r-1}  +  \tau^{r}\circ \mathcal{K}_{r} \tau^{r-1} \right] \tilde h_U =\tilde{\mathcal{h}}_{Z_1}\tau^{r}\circ \mathcal{K}_{r} \tau^{r-1}\tilde h_U .
\ee
The first term in the middle expression vanishes because $  \tau^{r-1} \tilde h_{U}$ is supported on $[U]_{D(r-1)}$ and hence is not supported on $Z_1$, so the commutator with $\tilde h_{Z_1}$ vanishes.

In a similar fashion,
\be
 \mathcal{K}_{r} \tau^{r-1}\tilde h_U 
 &= \mathcal{K}_{r} \left[ \tau^{r-2}  +  \tau^{r-1}\circ \mathcal{K}_{r-1} \tau^{r-2} \right] \tilde h_U = \mathcal{K}_{r} \tau^{r-1}\circ \mathcal{K}_{r-1} \tau^{r-2}\tilde h_U ,
\ee
since $  \tau^{r-2} \tilde h_{U}$ is supported on $[U]_{D(r-2)}$ and so its commutator with $\mathcal{K}_{r} $ vanishes.

Continuing, we get the  identity
\be\label{BY3}
\tilde{\mathcal{h}}_{Z_1} \tau^{r} \tilde h_U&  =
\tilde{\mathcal{h}}_{Z_1} \tau^{r} \circ {\mathcal K}_{r} \tau^{r-1}  \circ  \ldots \circ {\mathcal K}_{1} \tau^{0}  \tilde h_U .
\ee

We stress that, by construction, for any $j=1,\dots, r$,
\be\label{eq:Bj}
B_j=  \tau^{j}  \circ \mathcal{K}_{j}   \ldots \circ {\mathcal K}_{1} \tau^{0}  \tilde h_U
\ee
is supported on $[U]_{Dj}$. 

\subsection{Insertion of boson number cutoffs}
 For $j\le r$, we  define the annuli 
 \be
 S_{j}=\pa{[U]_{D(j+1)}\setminus [U]_{Dj}}\cap S.
 \ee
 We note that by construction, sets $S_k$ are disjoint and 
\be\label{deco-shell}
[U]_{rD}\cap S=(U\cap S)\cup\pa{\cup_{j=0}^{r-1} S_j}.
\ee
 We set
\be\label{pk}
\mathcal P_{j}=\chi_{N_{S_{j}}\le Mqr^{d-1}\ell^5 },\quad \bar{\mathcal P}_{j}=I- \mathcal P_{j}.
\ee
Unfortunately, we cannot demonstrate how the insertion of these projections work in the compact superoperator notation of \eqref{BY3}, so we rewrite it as an operator product first. Specifically,
we have 
\begin{multline}\label{Hamm}
\pa{{\mathcal K}_{r} \tau^{r-1}  \circ  \ldots \circ {\mathcal K}_{1} \tau^{0}  \tilde h_U }(s_r,s_0)\\ =\int_{G_r}\sum_{\sigma\in\set{0,1}^r}(-1)^{|\sigma|}\prod_{j=1}^r\pa{K_j^{1-\sigma_j}(s_j) U^{j-1}(s_{j},s_{j-1})}\,\tilde h_U(s_0)\, \prod_{j=1}^r\pa{ U^{j-1}(s_{j-1},s_{j})K_j^{\sigma_j}(s_{j})}d\vec s,
\end{multline}
where the left product is ordered from $r$ to $1$, the right product from $1$ to $r$, and subsequent products over $j$ run from larger to smaller indices. Here
\begin{itemize}
\item $|\sigma|$ is the Hamming weight of a string $\sigma$ (its $\ell_1$ norm) and $\sigma_j$ is its $j$-th component;
\item $U^j$ is the Schr\"odinger evolution (see \eqref{Schr}) generated by $ \tilde H_j$ defined in \eqref{H_j};
\item $K_j^{\alpha}$ stands for the formal $\alpha$-power of the operator $K_j$, i.e., $K_j^{0}=\mathds{1}$ and $K_j^{1}=K_j$.
\item $\vec s=(s_1,s_2,\ldots,s_{r-1})$ and $G_r$ is the simplex 
\be\label{simpl}
G_r=\left\{ \vec s \in \mathbb{R}^{r-1} \;\middle\vert{}\; s_0 \le s_1 \le \cdots \le s_{r-1} \le s_r \right\}.
\ee
\end{itemize}
Since every term in \eqref{Hamm} commutes with $\chi_{N_{\Theta}<  \xi}$ and $\norm{\tilde h_U}\le C_DQ$, we deduce that in order to estimate the norm of the expression in \eqref{Hamm} times $\chi_{N_{\Theta}<  \xi}$, it suffices to bound
\be\label{prodl}
\norm{\prod_{j=1}^r\pa{K_j^{1-\sigma_j}(s_j) U^{j-1}(s_{j},s_{j-1})}\chi_{N_{\Theta}<  \xi}}
\ee
for every $\sigma\in\set{0,1}^r$. 

We next decompose
\be
\prod_{j=1}^r\pa{K_j^{1-\sigma_j}(s_j) U^{j-1}(s_{j},s_{j-1})}\chi_{N_{\Theta}<  \xi}  =\sum_{\sharp} \prod_{j=1}^r\pa{\mathcal P_{j}^\sharp K_j^{1-\sigma_j}(s_j) U^{j-1}(s_{j},s_{j-1})}\chi_{N_{\Theta}<  \xi} ,
\ee
where $\mathcal P_{j}^\sharp$  stands  for either $\mathcal P_{j}$ or $\bar{\mathcal P}_{j}$.

Since $\mathcal P_{j}$ commutes with $B_j$ defined in \eqref{eq:Bj} we deduce that 
\be\label{BY2}
\prod_{j=1}^r\pa{K_j^{1-\sigma_j}(s_j) U^{j-1}(s_{j},s_{j-1})}\chi_{N_{\Theta}<  \xi}  =\sum_{\sharp} \prod_{j=1}^r\pa{\mathcal P_{j}^\sharp K_j^{1-\sigma_j}(s_j) U^{j-1}(s_{j},s_{j-1})} \prod_{j=1}^{r} \mathcal P_{j}^\sharp\, \chi_{N_{\Theta}<  \xi}.
\ee

\section{Bounding \texorpdfstring{$D_k$}{Dk} and completing the argument}
In this section we complete the argument by establishing the sharp estimate on $D_k$ and putting together the bounds we developed along the way. We will use the path expansion from the previous section with the choice $S=[Y]_{Dx_k}\cap [U]_{Dr}$). Fine tuning the various parameters (e.g., $Q, r$, and $L$) we will finally arrive at the desired bounds stated in Theorem \ref{kthm}.  In the argument below, the estimates focus on the forward evolution; to control  the backward evolution, we simply reverse time. 

Using \eqref{BY2},
\be\label{BY5}
\norm{\prod_{j=1}^r\pa{K_j^{1-\sigma_j}(s_j) U^{j-1}(s_{j},s_{j-1})}\chi_{N_{\Theta}<  \xi} }\le \sum_{\sharp} \norm{\prod_{j=1}^r\pa{\mathcal P_{j}^\sharp K_j^{1-\sigma_j}(s_j) U^{j-1}(s_{j},s_{j-1})} \prod_{j=1}^r \mathcal P_{j}^\sharp\, \chi_{N_{\Theta}<  \xi}}.
\ee
We recall that $\Theta=[U]_{D(r+1)}$, $\xi=qr^{d}M$ and set  $Q=\lceil qr^dM\ell^2\rceil$. We choose $\ell=\lceil\ln r\rceil $ for the parameter $\ell$ in \eqref{pk}. 
%
%

The number of $\bar{\mathcal P}_j$ projections in the non-vanishing product on the right hand side cannot exceed $r\ell^{-5}$, for otherwise 
$
N_{\Theta}>  qMr^d
$ 
since sets $ S_{j}$ are non-overlapping and satisfy \eqref{deco-shell}, 
and the corresponding contribution to $\sum_{\sharp}$ vanishes due to the presence of $\chi_{N_{\Theta}<  \xi} $ after  the product of projections. We will say that the index $j$ (which we will call a site $j$) is {\em bad} if  $\mathcal P_{j}^\sharp=\bar{\mathcal P}_{j}$ in  \eqref{BY5}. Otherwise, we will say that the site $j$ is {\em good}. 


We next aggregate good sites into discrete intervals $\set{I_m}=\set{[a_m,b_m]}_{m=1}^p$ and notice that since the number of bad sites is bounded by $r\ell^{-5}$, $p\le r\ell^{-5}+1$. For each $I_m$, we consider the operator
\be
B_m=
\prod_{j=a_m}^{b_m}\pa{\mathcal P_{j} K_j^{1-\sigma_j}(s_j) U^{j-1}(s_{j},s_{j-1})}.
\ee
Let $l_m=\lfloor(b_m-a_m+1)/\ell^2\rfloor$ and $S_i=[b_m-(i+1)\ell^2+1,b_m-i\ell^2]$.
We decompose $B_m$ as a product of blocks
\be
B_m=B_m^{(0)} B_m^{(1)} \ldots B_m^{(l_m)},
\ee
where for $0\le i<l_m$
\be
B_m^{(i)}=\prod_{j\in S_i}\pa{\mathcal P_{j} K_j^{1-\sigma_j}(s_j) U^{j-1}(s_{j},s_{j-1})}
\ee
and
\be
B_m^{(l_m)}=\prod^{b_m-l_m\ell^2}_{j=a_m}\pa{\mathcal P_{j} K_j^{1-\sigma_j}(s_j) U^{j-1}(s_{j},s_{j-1})}.
\ee
We note that
\be
B_m=\tilde B_m^{(0)}\tilde B_m^{(1)}\ldots\tilde B_m^{(l_m-1)}B_m^{(l_m)},
\ee
where
\be\label{hatBj1}
\tilde B_m^{(i)}&=\tilde B_m^{(i)}(s_{b_m-i\ell^2},s_{b_m-(i+1)\ell^2})=\prod_{j\in S_i}\pa{\mathcal P_{j} K_j^{1-\sigma_j}(s_j) U^{j-1}(s_{j},s_{j-1})}\prod_{j\in S_i}\mathcal P_j.
\ee
We are now ready to estimate the expression in \eqref{BY5}. For all $K_j$ that appear either in $B_m^{(l_m)}$ blocks or in bad sites, we use the rough estimate
\be\label{rough}
\norm{\mathcal P^\sharp K_j U }\le  \norm{K_j}=\norm{\tilde H_j- \tilde H_{j-1}}\le C_{d,D} \,Q \, r^d.
\ee
On the other hand, the remaining $K_j$ (located in the blocks $B_m^{(i)}$ with $0\le i<l_m$) will be bounded using the following estimate. To formulate it, we note that $\tilde B_m^{(i)}(s,s')$ is defined over the time interval with $s,s'\in[t_{k-1},t_k]$.
\begin{lemma}\label{block'}
For each $m$ and $0\le i<l_m$ we have
\be\label{eq:block}
\norm{\tilde B_m^{(i)}(s,s')\chi_{N_{\Theta}<  \xi}}\le \pa{CMqr^{d-1}\ell^7}^k\pa{CQr^d}^be^{C\ell^2|s-s'|},
\ee
where $k$ and $b$ satisfy $k+b=\sum_{j\in S_i}(1-\sigma_j)$ and $b\le2C_{d,D}\ell+2$.
\end{lemma}
We note that since all operators in $D_k$ commute with $\chi_{N_{\Theta}<  \xi}$, insertion of this projection into each $\tilde B_m^{(j)}$ is justified.

We now do a simple counting: The number of times we used the rough estimate in bounding each $K_j$ in the two operator products in \eqref{Hamm} cannot exceed the number of full blocks times $C\ell$ plus the number of intervals $I_m$ times $C\ell^2$ plus the number of bad sites. This yields an upper bound
\be
C\ell\frac r{\ell^2}+C\ell^2\pa{\frac r{\ell^5}+1}+C\frac r{\ell^5}
\le\frac{Cr}{\ell}
\ee
on the number of times we used the rough estimate \eqref{rough}, for sufficiently large $r$. On the other hand, the bound on each of the remaining $K_j$s 
 is essentially $CMqr^{d-1}\ell^7$ (we will account for the remaining factor $e^{C\ell^2|s-s'|}$ in \eqref{eq:block} at the final count stage).  Since $(1-\sigma_j)+\sigma_j=1$, there are exactly $r$ factors $K_j$ in the two products, and the additional factors from the rough estimates satisfy
\be
\frac{Qr^d}{Mqr^{d-1}\ell^7}\le Cr^{d+1},\qquad\pa{Cr^{d+1}}^{Cr/\ell}\le C^r.
\ee
Thus they yield the factor $\pa{CMqr^{d-1}\ell^7}^r$ in the estimation of \eqref{Hamm}. The sums over $\sigma$ and the two choices of projections contribute at most $8^r$, which is absorbed in $C^r$. For the right product we use its adjoint, since $K_j^*=-K_j$.

Taking into account the fact that the integration over the simplex in \eqref{Hamm}, together with the outer time integration in \eqref{BY3}, contributes the volume factor $(|t|(r+2)/L)^r/r!$, and that $\norm{\tilde{\mathcal h}_{Z_1}B}\le CQ\norm B$ and $\norm{\tilde h_U}\le CQ$, we deduce that for $t_{k-1}\le u\le s\le t_k$
\be\label{trestbmo2'}
\norm{D_k(u,s)\chi_{N_\Theta<\xi}}
&\le CQe^{C\ell^2|t|(r+2)/L}
\frac{\pa{CMqr^{d-1}\ell^7|t|(r+2)/L}^r}{r!}\norm{\tilde h_U}\\
&\le CQ^2e^{C\ell^2|t|(r+2)/L}
\frac{\pa{CMqr^{d-1}\ell^7|t|(r+2)/L}^r}{r!},
\ee
 where we used the fact that the product of all $e^{C\ell^2|s-s'|}$ factors originating from  \eqref{eq:block} cannot exceed $e^{C\ell^2|t_k-t_{k-1}|}=e^{C\ell^2|t|(r+2)/L}$.
We are now going to choose the parameter $L$. We choose $L$ to satisfy
\be\label{ellN}
\frac{L}{r+2}\ge CMq(1+|t|)r^{d-2}\ell^8.
\ee
Using \eqref{ellN} with a sufficiently large constant and $r!\ge(r/e)^r$, we conclude that
\be\label{trestbmo2'1}
e^{C\ell^2|t|(r+2)/L}\frac{\pa{CMqr^{d-1}\ell^7|t|(r+2)/L}^r}{r!}
\le\pa{\frac e\ell}^r,
\ee
where we also used $C\ell^2|t|(r+2)/L \le Cr/\ell^6$.

The first boundary sum in~\eqref{ges} has at most $C|Y|\langle L\rangle^d$ terms; the second has at most $Cr^d$ terms for each~$U$. By~\eqref{redb2},~\eqref{trestbmo2'}, and~\eqref{trestbmo2'1}, plugging this into~\eqref{decex'}, we get
\be
&\norm{\Pi_q\pa{\tau_{t,0}^{\tilde H}A_Y-\pa{\prod_{j=1}^{L/(r+2)}\tau_{t_j,t_{j-1}}^{\tilde H_j}}A_Y}\Pi_q}\\
&\le C\norm{A_Y}|Y|\langle L\rangle^d\frac{L}{r+2}\pa{\frac{|t|(r+2)}{L}}^2Q^2r^d\left[\pa{\frac e\ell}^r+e^{-cr^d}\right]\\
&\le C\norm{A_Y}|Y|\langle L\rangle^d(1+|t|)^2Q^2r^d\left[\pa{\frac e\ell}^r+e^{-cr^d}\right].
\ee
For the Hamiltonian truncation error~\eqref{trun1}, we choose $R=r$ and $V=\lceil C_{d,D}r^d\rceil$.  Since we use \eqref{constr} for the argument to work, we need $r\ge C_{d,D}(1+|t|)$. If $M$ is sufficiently large, then
\be
\norm{\Pi_q\pa{\tau_{t,0}^{H}A_Y-\tau_{t,0}^{\tilde H}A_Y}\Pi_q}
&\le C\norm{A_Y}|t||Y|\langle L\rangle^dQ\pa{\frac{C}{M\ell^2}}^V
\\&\le C\norm{A_Y}|t||Y|\langle L\rangle^dQe^{-cr^d}.
\ee
Combining these estimates with~\eqref{decex'} proves \eqref{eq:mbn}.  We note that  $A_Y(t)$ is a composition of local unitary conjugations, so $\norm{A_Y(t)}=\norm{A_Y}$.

 Recalling \eqref{ellN}, we can choose $L$ satisfying \eqref{eq:L}, completing the proof of Theorem~\ref{kthm}. \qed

\begin{proof}[Proof of Lemma \ref{block'}]
It will be convenient to temporary redefine the $\hat T$ operator introduced in \eqref{hatT}  as  
\be\label{hatT'}
\hat T=\sum_{x,y\in S} T(x,y)b^*_yb_x.
\ee
For a fixed indices $i,m$, let $Z$ be the annulus 
\be
Z=[U]_{D(b_m-i\ell^2-C_{d,D}\ell+1)}\setminus [U]_{D(b_m-(i+1)\ell^2+C_{d,D}\ell)}
\ee
with $C_{d,D}$ sufficiently large and $\ell>2C_{d,D}+2$, and for each $j\in S_i$ define $p_j=b_m-i\ell^2-j+1$.  Note that $h\mapsto p_h$ is a bijection from $S_i$ onto $\set{1,\dots,\ell^2}$. For $j\in S_i\cup\{\max S_i+1\}$, set
\be\label{eq:rang}
k_j&=\sum_{\substack{h\in S_i,\ h\ge j\\C_{d,D}\ell\le p_h<\ell^2-C_{d,D}\ell}}(1-\sigma_h)\in\{0,\ldots,\ell^2\},
\ee
 then 
\be k=k_{\min S_i} =\sum_{\substack{h\in S_i\\ C_{d,D}\ell\le p_h<\ell^2-C_{d,D}\ell}}(1-\sigma_h) \ee 
is exactly the number of indices with $\sigma_h=0$ in the middle region of $S_i$. Moreover, letting 
\be 
b=\sum_{h\in S_i}(1-\sigma_h)-k =\sum_{\substack{h\in S_i\\ p_h<C_{d,D}\ell\ \text{or}\ p_h\ge \ell^2-C\ell}} (1-\sigma_h),
\ee
we see that $b$ counts  $\sigma_h=0$ factors in the two boundary regions of $S_i$  and satisfies the condition stated in the lemma.

Let $\Psi_j=\hat T_Z^{eD}+\ell^2+p_j(p_j+1)$ and $\tilde\Psi_j=\hat T_Z^{eD}+\ell^2+(p_j+1)(p_j+2)$. 
We then can define the deformed operators $\hat K_{j}$ and $\hat U^{j-1}$ as follows:
\be\label{tildeop}
\hat K_{j}
=\begin{cases}
K_j,&p_j\in[1,C_{d,D}\ell),\\
\Psi_j^{k_{j+1}}K_j\Psi_j^{-1-k_{j+1}}
,&p_j\in[C_{d,D}\ell,\ell^2-C_{d,D}\ell),
\\
\Psi_j^{k}K_j\Psi_j^{-k}
, &p_j\in[\ell^2-C_{d,D}\ell,\ell^2]
\end{cases}
\ee
and
\be\label{tildeop1}
\hat U^{j-1}
=\begin{cases}
U^{j-1},&p_j\in[1,C_{d,D}\ell),\\
\Psi_j^{k_j}U^{j-1}\tilde \Psi_j^{-k_j}
,&p_j\in[C_{d,D}\ell,\ell^2-C_{d,D}\ell),\\
\Psi_j^{k}U^{j-1}\tilde\Psi_j^{-k}
,&p_j\in[\ell^2-C_{d,D}\ell,\ell^2].
\end{cases}
\ee
We can then rewrite $\tilde B_m^{(i)}$ in \eqref{hatBj1} as 
\be\label{hatBj'1}
\prod_{j\in S_i}\pa{\mathcal P_{j}^\sharp \hat K_j^{1-\sigma_j}(s_j) \hat U^{j-1}(s_{j},s_{j-1})}\times \pa{\hat T_Z^{eD}+\ell^2+(\ell^2+1)(\ell^2+2)}^{k}\prod_{j\in S_i}\mathcal P_{j}.
\ee

Furthermore, letting 
\be
\Pi_{\xi,\Theta}=\prod_{x\in\Theta} \chi_{n_x\le \xi}
\ee
and noting that
\be
\bar \Pi_{\xi,\Theta}\chi_{N_{\Theta}<  \xi} =0,
\ee
we can bound 
\be
\norm{\tilde B_m^{(i)}\chi_{N_{\Theta}<  \xi}}\le \norm{\tilde B_m^{(i)}\Pi_{\xi,\Theta}}.
\ee

Using Corollary \ref{cor} (applied to the adjoint propagator), Lemma \ref{lem1a}   with $R=eD$ (note that the same statements hold for the redefined $\hat T$ operator in this context), and
\be
(p_j+1)(p_j+2)-p_j(p_j+1)=2p_j+2
\ge2k_j+1,
\ee
we can bound $\norm{\hat U^{j-1}(s_j,s_{j-1})}\le e^{C|s_j-s_{j-1}|\ell^2}$, 
\be
\norm{\hat K_j}\le\begin{cases}
C,&p_j\in[C_{d,D}\ell,\ell^2-C_{d,D}\ell),\\
CQr^d,&p_j\in[1,C_{d,D}\ell)\cup[\ell^2-C_{d,D}\ell,\ell^2].
\end{cases}
\ee
Here $\supp K_j\subset Z$ in the first case, so $[\hat T_Z^{eD},K_j]=0$; in the other cases the conjugation exponent is at most $\ell^2$ and the shift is at least $\ell^2$.

This leads to the estimate
\be
&\norm{\tilde B_m^{(i)}(s,s')\chi_{N_{\Theta}<  \xi}}\le C^{k}
(CQr^d)^{b}e^{C|s-s'|\ell^2}\norm{\pa{\hat T_Z^{eD}+\ell^2+(\ell^2+1)(\ell^2+2)}^{k}\prod_{j\in S_i}\mathcal P_j\Pi_{\xi,\Theta}}.
\ee
It follows from the definitions of $\mathcal P_{k}$ and $\Pi_{\xi,\Theta}$  that
\be\label{fesq'}
\norm{\pa{\hat T_Z^{eD}+\ell^2+(\ell^2+1)(\ell^2+2)}^{k}\prod_{j\in S_i}\mathcal P_j\Pi_{\xi,\Theta}}
&\le\pa{CMqr^{d-1}\ell^7+\xi|S|e^{-(C_{d,D}\ell-1)/e}+C\ell^4}^{k}\\
&\le\pa{CqMr^{d-1}\ell^7}^{k},
\ee
for  $M$ large enough, where we used $|S|\le C_{d,D}r^d$ and $e^{-\ell}\le1/r$. The result  follows since there are at most $2C_{d,D}\ell+2$ indices outside the middle region of $S_i$ and $Qr^d/(Mqr^{d-1}\ell^7)\le Cr^{d+1}$.
\end{proof}

\section{Proof of Corollary \ref{thm}}
We will that for an observable $A_Y(t)$ supported on $[Y]_{DL}$ with $L$ as in \eqref{eq:r''} we have the bound
\be\label{trnor}
\norm{ \rho^{1/2} \pa{\tau_{t,0}^{H}A_Y- A_Y(t)}\rho^{1/2}}_1 \le C_p\norm{A_Y}\abs{Y}|t|^{-p},
\ee
where $\norm{\cdot}_1$ stands for the trace norm.

For $q\in\N$, $\kappa>0$, and $x\in\mathcal V$, let 
\[f(x,q,\kappa)=q\pa{\kappa^2\ln\pa{d\pa{x,Y}+1}+1}^{\frac1\kappa}.\]
 We set
\be
\Pi_{q,\kappa}(X)=\prod_{x\in X}\chi_{n_x\le f(x,q,\kappa)}
\ee
and let $\bar \Pi_{q,\kappa}(X)=\mathds{1}-\Pi_{q,\kappa}(X)$. For $\kappa=0$, we set
\be
\Pi_{q,0}(X)=\Pi_{q}(X).
\ee
We observe that, denoting $\mu_R= \pa{\kappa^2\ln\pa{R+1}+1}^{\frac1\kappa}$, for any observable $A$ supported on $[Y]_R(X)$ 
\be\label{pro}
\norm{A \Pi_{q,\kappa}(X)}\le \norm{A \Pi_{q\mu_R}(X)}.
\ee
For any observable $B$, we have
\be
\norm{\rho^{1/2} B \rho^{1/2}}_1&\le \norm{\rho^{1/2} \Pi_{q,\kappa}(X) B \rho^{1/2}}_1+\norm{\rho^{1/2} \bar \Pi_{q,\kappa}(X) B \rho^{1/2}}_1\\&\le  \norm{\rho^{1/2} \Pi_{q,\kappa}(X) B \rho^{1/2}}_1+\norm{\rho^{1/2} \bar \Pi_{q,\kappa}(X)}_{HS} \norm{B} \\&\le \norm{\rho^{1/2} \Pi_{q,\kappa}(X) B \Pi_{q,\kappa}(X) \rho^{1/2}}_1+2\norm{\rho^{1/2} \bar \Pi_{q,\kappa}(X)}_{HS} \norm{B} \\&\le \norm{ \Pi_{q,\kappa}(X) B \Pi_{q,\kappa}(X)}+2\norm{\rho^{1/2} \bar \Pi_{q,\kappa}(X)}_{HS} \norm{B} .
\ee
 Here $\|\cdot\|_{HS}$ stands for the Hilbert-Schmidt norm. 
Since 
\[
0\le \bar \Pi_{q,\kappa}(X)\le \sum_{x\in X}\chi_{n_x> f(x,q,\kappa)},
\]
with the aid of \eqref{eq:locde} we can bound 
\be
\norm{\rho^{1/2} \bar \Pi_{q,\kappa}}^2_{HS} 
&= \tr \rho\bar \Pi_{q,\kappa}(X)\le \sum_{x\in X}
\norm{e^{-(n_x)^\kappa} \chi_{n_x> f(x,q,\kappa)}}\tr \pa{\rho e^{(n_x)^\kappa}}
\\ &\le Ce^{-q^\kappa}\sum_{x\in X}\pa{1+d\pa{x,Y}}^{-\kappa^2 q^\kappa}\le C_{\kappa}e^{-q^\kappa}\abs{Y}
\ee
as long as  $q> q_0=q_0(\kappa)$ with $\kappa^2 q_0^{\kappa}=d+1$. 
 Using it with $B=\tau_{t,0}^{H}A_Y-A_Y(t)$ from \eqref{kthm},  we get
\be\label{pa21}
&\norm{ \rho^{1/2} \pa{\tau_{t,0}^{H}A_Y- A_Y(t)}\rho^{1/2}}_1 \le \norm{\Pi_{q,\kappa}(\mathcal V)\pa{\tau_{t,0}^{H}A_Y-A_Y(t)}\Pi_{q,\kappa}(\mathcal V)}+C_{\kappa}e^{-q^\kappa/2}\abs{Y}\norm{A_Y}.
\ee 
We estimate $\norm{\Pi_{q,\kappa}(\mathcal V)\pa{\tau_{t,0}^{H}A_Y-A_Y(t)}\Pi_{q,\kappa}(\mathcal V)}$ by slightly modifying the proof of Theorem \ref{kthm}. For $Z\subset[Y]_{C_DL}$ and $1\le R\le C_DL$, we have
\be\label{R}
\norm{\hat T_Z^R\Pi_{q,\kappa}(\mathcal V)}&\le q\sum_xe^{-d(x,Z)/R}\pa{1+\kappa^2\ln(1+d(x,Y))}^{1/\kappa}\\&\le C_\kappa q\mu_L\sum_xe^{-d(x,Z)/(2R)}.
\ee
For the choices of our parameters in the truncation and high-number estimates, the last sum is at most $V=\lceil C_{d,D}r^d\rceil$. Specifically, \eqref{smb1} is replaced by
\be\label{smb'}
\norm{\Pi_{q,\kappa}( \mathcal V)\tau_{t,s}^{H'}\pa{\hat T_Z^R+l}^V}\le 
 \pa{C_\kappa q\mu_LV}^V,\quad 1\le l\le4V,\quad 4eD|t-s|\le R,
\ee
while \eqref{trun1} is replaced by
\be\label{trun1'}
\norm{\Pi_{ q,\kappa }( \mathcal V)\pa{ \tau_{t,0}^{H'}A-\tau_{t,0}^{\tilde H'}A}\Pi_{ q,\kappa }( \mathcal V)}\le 
C_{d,D}\norm{A}|t|  \abs{ [Y]_{D\langle L\rangle}}\pa{C_\kappa q\mu_LV}^VQ^{1-V}.
\ee
Following the rest of the argument, with $C_\kappa q\mu_L$ in place of $q$ in $\xi,Q$, we obtain
\be
\norm{\Pi_{q,\kappa}(\mathcal V)\pa{\tau_{t,0}^{H}A_Y-A_Y(t)}\Pi_{q,\kappa}(\mathcal V)}&\le C\norm{A_Y}\abs{Y}\langle L\rangle^d(1+|t|)^2Q^2r^d\left[\pa{\frac e\ell}^r+e^{-cr^d}\right],
\ee 
provided $r\ge C(1+|t|)$ and $\lfloor L/(r+2)\rfloor\ge CMq\mu_L(1+|t|)r^{d-2}\ell^8$, using radius $(r+2)\lfloor L/(r+2)\rfloor\le L$ in the construction.

We choose
\be
q=\left\lceil\max\left\{q_0+1,\pa{2(p+1)\ln|t|}^{1/\kappa}\right\}\right\rceil.
\ee
We set
\be
r=\lceil C(1+|t|)\rceil.
\ee
Then $e^{-q^\kappa/2}\le |t|^{-p-1}$ and, with $L$ as in \eqref{eq:r''}, $q\mu_L\le C_p(\ln|t|)^{2/\kappa}$ and 
\[\lfloor L/(r+2)\rfloor\ge CMq\mu_L(1+|t|)r^{d-2}\ell^8\]
 for sufficiently large $C_p$. Moreover,
\be\label{eq:penu}
\langle L\rangle^d(1+|t|)^2Q^2r^d\left[\pa{\frac e\ell}^r+e^{-cr^d}\right]\le C_p|t|^{-p},
\ee
since the prefactor is polynomial in $|t|$ and $\ln|t|$ while $r\ge c|t|$. Plugging it into \eqref{pa21} yields \eqref{trnor}.
\qed

\appendix 
\section{Particle transport bounds for long-range systems}\label{app:A_long_range}
 We will assume that instead of the finite-range hopping assumption and \eqref{eq:J}, for some $\alpha>d$ the following condition holds:
\be\label{eq:J'}
J:=\sup_t\sup_{x,y\in\mathcal V}\abs{J_{\langle x,y\rangle}(t)}\langle d(x,y)\rangle^{\alpha}<\infty.
\ee

Let $\beta>0$ satisfy  
\be\label{beta}
\begin{cases}\beta<\alpha-d & \alpha\le d+1\\ \beta=1&\alpha>d+1\end{cases}
\ee

Th following result is the long-range counterpart to Proposition \ref{nrm1}.
\begin{proposition}
\label{nrm1'}
Let $H'$ be of the form \eqref{H'}, and fix $q,R\in\N$ with $q>3$ and $Z\subset \mathcal V$. Then, for for any $l\in\N$ and $R>5$ that satisfies $C(\alpha-\beta,d)J\abs{Z}|t-r|\le R^\beta$, $l\le \abs{[Z]_R}$, there exists a constant $K_d=K_d(\alpha-\beta)\ge1$ such that  
\be\label{smb1'}
\norm{\Pi_{ q }( \mathcal V)\tau_{t,r}^{H'} \pa{\hat T_{Z}^R+l}^{V}}\le \pa{e^3qK_d V}^{V+\frac l 2},\quad V=\lceil C(\alpha-\beta,d)\abs{Z}R^d\rceil.
\ee
\end{proposition}

\subsection{Proof}

The analog of Lemma \ref{lem:T} is

 \begin{lemma}
 Let $R\ge 1$, $\alpha>d$, $Z\subset\mathcal V$,  and $a\ge 1 $.  Define $T_{Z}^R$ as a self-adjoint multiplication  operator on $\ell_2(G)$ with
 \be
T_{Z}^R(x,x)=\langle d(x,Z)/R\rangle^{\beta-\alpha}. 
 \ee
 Then for any odd $k\in\N$, any $H'$, and any $Z\subset\mathcal V$ the following operator inequality holds:
  \be
 i\left[ H'(t), \pa{\hat T_{Z}^R+a}^k\right]\le   \frac{C(\alpha-\beta,d)J\abs{Z}k}{R^\beta}\pa{\hat T_{Z}^R+a'}^{k},\quad a'=a+1.
 \ee

 \end{lemma}
\begin{proof}
The argument follows closely the one for Lemma \ref{lem:T}. The $K$ in \eqref{eqK} is replaced with 
\be\label{eqK'}
K(x,y)=-i\pa{\langle d(x,Z)/R\rangle^{\beta-\alpha}-\langle d(y,Z)/R\rangle^{\beta-\alpha}}J_{\langle x,y\rangle}.
\ee
Lemma \ref{kerest} is modified to 
\begin{lemma}\label{kerest'}
For $\alpha>d$ and $\beta$ as in \eqref{beta}, 
\be
\sum_{y\in\mathcal V}\langle d(x,y)\rangle^{-\alpha}\abs{\langle d(x,Z)/R\rangle^{\beta-\alpha}-\langle d(y,Z)/R\rangle^{\beta-\alpha}}\le \frac{C(\alpha-\beta,d)}{\displaystyle R^{\beta}}\abs{Z}\langle d(x,Z)/R\rangle^{\beta-\alpha}.
\ee
\end{lemma}
This yields the estimates
\be\label{eqKxy'} 
\sum_{y\in  \mathcal V}  \abs{K(x,y)}&\le \frac{C(\alpha-\beta,d)J\abs{Z}}{\displaystyle R^{\beta}}\langle d(x,Z)/R\rangle^{\beta-\alpha},\\ \sum_{x\in  \mathcal V}  \abs{K(x,y)}&\le \frac{C(\alpha-\beta,d)J\abs{Z}}{\displaystyle R^{\beta}}\langle d(y,Z)/R\rangle^{\beta-\alpha}
\ee 
As a result, instead of \eqref{hatK1} we get
\be\label{hatK1'}
 i[H', \hat T_{Z}^R+a]&\le\frac{C(\alpha-\beta,d)J\abs{Z}}{\displaystyle R^{\beta}}\pa{\hat T_{Z}^R+a'}. 
\ee
The rest of the argument is identical to the one in Lemma \ref{lem:T}.
\end{proof}
\begin{proof}[Proof of Lemma \ref{kerest'}] 
For any $m>0$, we have the inequality (a consequence of the mean value theorem for the function $f(s)=\langle s\rangle^{-m}$)
\be
\left|\langle d(x,Z)/R\rangle ^{-m}-\langle d(y,Z)/R\rangle ^{-m}\right|&\le \frac{m}{R}d(x,y)\pa{\langle d(x,Z)/R\rangle ^{-m-1}+\langle d(y,Z)/R\rangle ^{-m-1}}\\&\le \frac{m}{R}d(x,y)\pa{\langle d(x,Z)/R\rangle ^{-m}+\langle d(y,Z)/R\rangle ^{-m}}.
\ee
Since trivially
\be
\left|\langle d(x,Z)/R\rangle ^{-m}-\langle d(y,Z)/R\rangle ^{-m}\right|\le\left|\langle d(x,Z)/R\rangle ^{-m}+\langle d(y,Z)/R\rangle ^{-m}\right|,
\ee
we can interpolate between these two bounds to obtain that 
\be\left|\langle d(x,Z)/R\rangle ^{\beta-\alpha}-\langle d(y,Z)/R\rangle ^{\beta-\alpha}\right|&\le \frac{\pa{\alpha-\beta}^\beta}{\displaystyle R^{\beta}}d^{ \beta}(x,y)\pa{\langle d(x,Z)/R\rangle  ^{\beta-\alpha}+\langle d(y,Z)/R\rangle^{\beta-\alpha}}.
\ee
Since 
\be
\langle d(x,Z)/R\rangle^{\beta-\alpha}\sum_{y\in\mathcal V}\langle d(x,y)\rangle^{\beta-\alpha}\le C(\alpha-\beta,d)\langle d(x,Z)/R\rangle ^{\beta-\alpha},
\ee
we only need to show the bound
\be\label{triC}
\sum_{y\in\mathcal V}\langle d(x,y)\rangle^{\beta-\alpha}\langle d(y,Z)/R\rangle^{\beta-\alpha}\le C(\alpha-\beta,d)\abs{Z}\langle d(x,Z)/R\rangle^{\beta-\alpha}.
\ee
Clearly, the inequality holds for $d(x,Z)<R$, so without loss of generality we will assume that $d(x,Z)\ge R$.

To this end, we consider the following  regions $A,B$ that cover $\mathcal V$, and estimate the summation separately:
\be
A=\set{y\in\mathcal V:\ d(y,Z)\le \frac12d(x,Z)}, \quad B=\set{y\in\mathcal V:\ d(y,Z)> \frac12d(x,Z)}.
\ee
For $y\in A$ we have $d(x,y)\geq \frac{1}{2}d(x,Z)$, which implies
\be
\sum_{y\in A}\langle d(x,y)\rangle^{\beta-\alpha}\langle d(y,Z)/R\rangle^{\beta-\alpha}&\le 2^{\alpha-\beta}R^{\alpha-\beta}d^{\beta-\alpha}(x,Z)\sum_{y\in A}\langle d(y,Z)\rangle^{\beta-\alpha}\\&\hspace{-3cm}\le C(\alpha-\beta,d) \abs{Z}R^{\alpha-\beta}d^{\beta-\alpha}(x,Z)\le C(\alpha-\beta,d) \abs{Z}d\langle d(x,Z)/R\rangle^{\beta-\alpha},
\ee
where in the last step we used $d(x,Z)\ge R$. 

For $y\in B$ we have $d(y,Z)> \frac{1}{2}d(x,Z)$, which implies
\be
\sum_{y\in B}\langle d(x,y)\rangle^{\beta-\alpha}\langle d(y,Z)/R\rangle^{\beta-\alpha}&\le 2^{\alpha-\beta}\langle d(x,Z)/R\rangle ^{\beta-\alpha}\sum_{y\in B}\langle d(x,y)\rangle^{\beta-\alpha}\\&\le C(\alpha-\beta,d) \langle d(x,Z)/R\rangle^{\beta-\alpha}.
\ee
\end{proof}
The argument identical to the one used in the proof of Corollary \ref{cor} now yields the following result.
\begin{corollary}\label{cor'} Let $R\ge 1$, then for any odd $k\in\N$ and any $t,r$, $l\ge1$ the following operator inequalities hold:
\be\label{comexpz1'}
\tau_{t,r}^{H'}\pa{\pa{\hat T_{Z}^R+l}^k}\le  \exp\pa{\frac {C(\alpha-\beta,d)J\abs{Z}|t-r|k}{R^\beta}}\pa{\hat T_{Z}^R+l+k}^{k}.
\ee
\end{corollary}
The proof of Proposition \ref{nrm1'} is then analogous to the proof of Proposition \ref{nrm1}.

\section{Basic properties of evolutions}\label{app:B}.

We  recall some basic properties of Schr\"odinger and Heisenberg dynamics, see \cite[Section X.12]{RS} for details and proofs. Let  $H(s)$ be a bounded self-adjoint operator, and let $U(t,s)$ be the corresponding Schr\"odinger evolution defined in \eqref{Schr}. Then
\begin{enumerate}
\item Norm preservation: $\norm{U(t,s)\psi}=\norm{\psi}$ for any $\psi\in\mathcal F$.
\item The semigroup property: $U(t,s)U(s,r)=U(t,r)$.
\item Time derivative in the second variable:
  \be
 -i\partial_{s}U(t,s)=U(t,s)H(s).
 \ee
 \item Direct sum decomposition:  If $H_X(t)$ commutes with the total number operator $N_X$, then $U_X(t,s)$ can be decomposed as a direct sum of unitaries generated by the restriction of $H_X(t)$ to  sectors $N_X=N$ with $N\in\N$. 
\end{enumerate}

These features induce  the following properties of $\tau_{t,s}$, defined in \eqref{Heis}: 
\begin{enumerate}
\item Connection to $U$: We have $\tau_{t,s}^{H} A=U(t,s)AU^*(t,s)=U(t,s)AU(s,t)$.
\item $\tau^{H}$ preserves the norm: $\norm{\tau_{t,s}^{H}A}=\norm{A}$.
\item Multiplicative property:  $\tau_{t,s}^{H} (AB)=\pa{\tau_{t,s}^{H} A} \pa{\tau_{t,s}^{H}B}$ for any $A,B$.
\item The analog of the semigroup property: $\tau_{t,r}^{H}\tau_{r,s}^{H} A=\tau_{t,s}^{H} A$.
\item Time derivative in the second variable:
  \be
 \partial_{s}\tau_{t,s}^{H} A=-\tau_{t,s}^{H}\mathcal H(s)A.
 \ee
 \item Decoupling property: If $H(t)=H_X(t)+H_Y(t)$, where $X$ and $Y$ are supports of $H_X(t)$ and $H_Y(t)$ respectively, and $X\cap Y=\emptyset$, $A$ is supported on $X$, then \[\tau_{t,s}^{H}A=\tau_{t,s}^{H_X}A.\]
\end{enumerate}

\section*{AI declaration}
This work was done without  AI except that AI (ChatGPT-6 Astra) was used to check the manuscript for errors, which led to minor corrections.

\section*{Acknowledgments}
A.E. was  supported in part by the NSF under grant DMS-2307093. The research of M.L.~is supported by the Deutsche Forschungsgemeinschaft (DFG, German Research Foundation) through grants TRR 352 – Project-ID 470903074 and FOR 5413 - Project-ID 465199066 and by the European Union (ERC StG MathQuantProp, Grant Agreement 101163620).%
\footnote{%
    Views and opinions expressed are however those of the authors only and do not necessarily reflect those of the European Union or the European Research Council Executive Agency.
    Neither the European Union nor the granting authority can be held responsible for them.
}

We are grateful to Andreas Deuchert for helpful discussions. 
\bibliographystyle{plain}
\bibliography{bibfile.bib}
\end{document}